\documentclass[]{aastex}

\usepackage{emulateapj5}
\usepackage{graphicx} 
\usepackage{ulem}
\usepackage{rotating}
\usepackage{lscape}

\def\simlt{$_<\atop{^\sim}$}
\def\simgt{$_>\atop{^\sim}$}
\def\q0{q$_0$}
\def\h0{H$_0$}
\def\eg{e.g.~}

\def\kms{km~s$^{-1}$ }

\def\cm2{cm$^{-2}$}
\def\nhi{\mbox{$N_{\rm HI}$}}
\def\atomcm{atoms cm$^{-2}$}
\def\etal{et~al.}
\def\lya{Ly$\alpha$}
\def\lyb{Ly$\beta$}
\def\civ{\ion{C}{4}}
\def\c2{\ion{C}{2}}
\def\o1{\ion{O}{1}}
\def\mgii{\ion{Mg}{2}}
\def\fe2{\ion{Fe}{2}}
\def\mg1{\ion{Mg}{1}}
\def\sil2{\ion{Si}{2}}
\def\si4{\ion{Si}{4}}
\def\al2{\ion{Al}{2}}
\def\n5{\ion{N}{5}}

\def\nd{\nodata}

\begin{document}

\title{Absorption Systems in the Spectra of sixty-six z $\ga$ 4 Quasars}

\author{C\'eline P\'eroux\footnotemark[1]} \affil{Institute of
Astronomy, Madingley Road, Cambridge CB3 OHA, UK}

\author{ Lisa J. Storrie-Lombardi} \affil{SIRTF Science Center,
California Institute of Technology, MS 100-22, Pasadena CA 91125, USA}

\author{Richard G. McMahon, Mike Irwin}
\affil{Institute of Astronomy, Madingley Road, Cambridge CB3 OHA, UK}

\and
\author{Isobel M. Hook} \affil{Institute for Astronomy, Royal
Observatory, Blackford Hill, Edinburgh EH9 3HJ, UK}

\footnotetext[1]{e-mail: celine@ast.cam.ac.uk}

\submitted{Accepted by the Astronomical Journal}

\begin{abstract}
We present high signal-to-noise, $\sim$ 5 \AA\ resolution (FWHM)
spectra of 66 $z \ga 4$ bright quasars obtained with the 4 m Cerro
Tololo Inter-American Observatory and 4.2 m William Hershel
telescopes. The primary goal of these observations was to undertake a
new survey for intervening absorption systems detected in the spectra
of background quasars. We look for both Lyman-limit systems (column
densities $N_{HI} \geq 1.6 \times 10^{17}$ atoms cm$^{-2}$) and damped
Ly$\alpha$ systems (column densities $N_{HI} \geq 2 \times 10^{20}$
atoms cm$^{-2}$).  This work resulted in the discovery of 49
Lyman-limit systems , 15 of which are within 3000 km s$^{-1}$ of the
quasar emission and thus might be associated with the quasar itself,
26 new damped Ly$\alpha$ absorption candidates, 15 of which have
$z>3.5$ and numerous metal absorption systems. In addition ten of the
quasars presented here exhibit intrinsic broad absorption lines.

\end{abstract}


\section{Introduction}

Because they are extremely luminous, quasars are among the youngest
objects observed in the Universe, and have now been detected out to
redshifts of $z \sim 5.8$ (Fan \etal\ 2000). Observing quasars at
such high redshifts gives us direct indications of the ionization
state of the early Universe. Indeed the lack of an observed
Gunn-Peterson effect (Gunn \& Peterson 1965) indicates that the
Universe is already ionized at these early epochs.

In addition to being important observations in their own right,
spectroscopic studies of quasars allow the detection of much fainter
systems, observed in absorption in the quasar spectra. There are two
main classes of quasar absorption lines: the metal systems, such as
CIV or MgII, and the more numerous hydrogen lines. The latter are
subdivided into three different groups according to their neutral
hydrogen column densities. The Lyman-$\alpha$ forest absorbers have
column densities $10^{12} - 1.6 \times 10^{17}$ atoms cm$^{-2}$. The
Lyman-limit systems (hereafter LLS) have $N_{HI} \geq 1.6 \times
10^{17}$ atoms cm$^{-2}$ and the damped Lyman-$\alpha$ systems
(hereafter DLA) are a subset of the LLS with $N_{HI} \geq 2 \times
10^{20}$ atoms cm$^{-2}$. They thus probe media with different column
densities spanning the range from voids through to halos and disks of
both dwarf and normal (proto)galaxies. The morphology of DLAs in
particular is still open to debate. The hypotheses put forward range
through large disk systems (Prochaska \& Wolfe 1998), low surface
brightness galaxies (Jimenez \etal\ 1999) and dwarf galaxies
(Matteucci \etal\ 1997).  Although the exact nature of the quasar
absorbers is not known, they form a sample of systems unbiased as
regards luminosity, specific morphology, metallicity, or emission line
strength, thus enabling studies of metallicity and HI evolution over a
large redshift range.

The primary goal of our spectroscopic campaign is to obtain a
statistically significant number of high-redshift absorbers to answer
the questions raised by the apparent deficit of high column density
systems at such redshifts (Storrie-Lombardi \etal\ 1996a;
Storrie-Lombardi \& Wolfe 2000).  In particular we aim to study in
more detail the evolution with redshift of the column density
distribution, number density, and comoving mass density of high column
density HI absorption systems.  The latter has been quite
controversial in the last few years. Lanzetta \etal\ 1995 found
that $\Omega_{DLA}$ at $z \sim 3.5$ is twice the value at $z \sim 2$,
implying a larger star formation rate than indicated by metallicity
studies. This created the so-called ``cosmic G-dwarf
problem''. Storrie-Lombardi \etal\ 1996b used new data to show that
$\Omega_{DLA}$ decreases at high-redshift thus solving the ``cosmic
G-dwarf problem''. The work of Storrie-Lombardi \& Wolfe 2000
confirmed such results by using a compilation of data gathered from
the litterature together with new spectroscopic observations. The
aim of our new survey for quasar absorbers is to better understand the
high-redshift end of the mass density of neutral hydrogen by
significantly improving the statistics at $z \ga 3.5$. At
low-redshift, recent studies by Rao \& Turnshek 2000 show that
$\Omega_{DLA}$ might be higher than first expected. Their analysis is
based on HST observations of quasars with knowm MgII systems (MgII
always being associated with DLA systems, the reverse not being
true). Nevertheless it should be emphasized, as the
authors themselves pointed at, that the analysis is based on a
relatively small number of systems. At all redshifts such measurements
can be used to constrain the most recent semi-analytical models of
galaxy formation (Kauffmann \& Haehnelt 2000 and Somerville \etal\
2000, which build on models originally presented by Kauffmann, White,
$\&$ Guiderdoni 1993 and Cole \etal\ 1994). Several fundamental
questions remain including locating the epoch of DLAs assembly,
clarifying the the relationship between Lyman limit systems and damped
absorbers, measuring the total amount of neutral hydrogen contained in
quasar absorbers and studying how this varies with redshift.  We will
discuss the impact of our new survey on these issues in more detail in
future papers currently in preparation (Peroux \etal\ 2000).

Any survey for quasar absorbers begins with a search for bright
quasars and so constitutes an ambitious observational program. We
observed sixty-six z \simgt 4 quasars discovered by various groups
(Storrie-Lombardi \etal\ 2000, Fan \etal\ 1999, Warren, Hewett \&
Osmer 1991, Kennefick \etal\ 1995a, Storrie-Lombardi \etal\ 1996a,
Zickgraf \etal\ 1997, Kennefick \etal\ 1995b, Henry \etal\ 1994, Hook
\etal\ in preparation and Hall \etal\ 1996) almost all of which have
not been previously studied at such resolution ($\approx$ 5 \AA) and
signal-to-noise (ranging from 10--30). We obtained optical spectra at
the 4.2 m William Herschel Telescope for the northern quasars and at
the 4 m Cerro Tololo Inter-American Observatory for the southern
objects. More information about $z \ga 4$ quasars is available at the
following URL: http://www.ast.cam.ac.uk/$\sim$quasars.

This paper is organized as follows. In \S 2 we provide the details 
of the set-up for each observational run and in \S 3 describe the data
reduction and present the quasar spectra. Accurate quasar
redshift and magnitude measurements are given in \S
4. The surveys for Lyman-limit systems and damped Ly-$\alpha$ 
absorbers are presented in \S 5 and \S 6, respectively. 
Provisional interpretation of metal absorption features 
are summarized in \S 7.  Notes on individual objects are provided in 
\S 8 and conclusions are presented in section \S 9.

\section{Observations}

All the new observations were carried out during two observing runs at
the 4.2 m William Herschel telescope (WHT) of the Isaac Newton Group
of telescopes in the Canary Islands and two observing runs at the
Blanco 4 m telescope at the Cerro Tololo Inter-American Observatory
(CTIO) in Chile.  High signal-to-noise optical spectrophotometry was
obtained covering approximately 3500 \AA\space to 9000 \AA, the exact
range depending on which instrument was used for the observations.  A
journal of the observations is presented in Table~\ref{t_obs_log}.

Thirty-one (including a z=1.90 quasar PSS J0052$+$2405) quasars were
observed at the WHT during 1998 September 22-24 and 1999 March
18-19. The integration times were typically 1800 -- 3600 seconds. We
used the ISIS double-spectrograph which consists of two independant
arms fed via a dichroic allowing for blue and red observations to be
carried out simultaneously. Gratings with 158 lines mm$^{-1}$ and a
dichroic to split the light at $\sim$5700 \AA\space were used. This
gives a dispersion of 2.89 \AA\space pixel$^{-1}$ in the red arm and
1.62 \AA\space pixel$^{-1}$ in the blue. The gratings were arranged so
that the blue part of the spectrum was centered on 4500 \AA\ while the
red was centered on 7000 \AA. On the blue arm a thinned coated English
Electric Valve (EEV) $2148 \times 4200$ CCD with 13.5 $\mu$m pixels
was used as detector. On the red arm a thinned coated Tektronix $1124
\times 1124$ CCD with 24 $\mu$m pixels was used. All the narrow-slit
observations were taken with a slit width of 1.2 -- 1.5 arcsec while
the wide-slit observations were carried out with a slit width of 5 --
7 arcsec.  Blind-offsetting from bright $\sim 15 $--$ 17^{th}$
magnitude stellar fiducials was used to position the quasars in the
slit partly to save acquisition time and partly because the majority
of the quasars were not visible using the blue sensitive TV
acquisition system. Readout time was reduced by windowing the CCDs in
the spatial direction.

Thirty-five quasars were observed at CTIO during 1998 October 14 --
16, and 1999 October 9 -- 12. The typical exposure time was 3600
seconds for the brighter objects (R$\sim 18 $--$ 19^{th}$ mag) but
substantially longer times were used for the fainter Sloan Digital Sky
Survey quasars. We used the R-C spectrograph with the 316 lines
mm$^{-1}$ grating, centered at 6050 \AA\ and covering the range
3000\AA $--$ 9100\AA. This set-up resulted in a dispersion of 1.98 \AA\
pixel$^{-1}$. The detector used was a Loral $3072 \times 1024$ CCD
detector. The narrow and wide slit observations were taken with 1 --
1.5 arcsec and 5 arcsec widths, respectively. Because of the
substantial wavelength coverage available with this set-up we used a
WG345 blocking filter (with 50 $\%$ transmission at 3450 \AA) to
minimize the second order contamination from the standard stars above
7000 \AA. The contamination is negligible for the quasars as most have
no flux below 4500 \AA\ but affects the standard stars that have
substantial flux at 3500 \AA. Appropriate measures, as discussed in
the data reduction section, have been taken so that this set-up does
not modify the flux calibration at the red end of the spectra. Using
two instrumental set-ups in order to completely remove the second
order contamination problem would have resulted in a 60--80 $\%$
increase in the required observing time.

The observations of SDSS~J0338+0022 were taken at the Keck Observatory
with the Low Resolution Imaging Spectrometer (LRIS).  The
observational details are given in Songaila \etal\ (1999).

\section{Data Reduction}
The data reduction was undertaken using the IRAF~\footnote{IRAF is
distributed by the National Optical Astronomy Observatories, which are
operated by the Association of Universities for Research in Astronomy,
Inc., under cooperative agreement with the National Science
Foundation.} software package. Because the bias frames for each nights were so
similar, a master `zero' frame for each run was created using the
IMCOMBINE routine.  The data were overscan corrected, zero corrected,
and trimmed using CCDPROC.  Similarly a single flat-field frame was
produced by taking the median of the Tungsten flats. The overall
background variation across this frame was removed to produce an image
to correct for the pixel-to-pixel sensitivity variation of the data.
The task APALL was used to extract 1-D multi-spectra from the 2-D
frames. The routine estimates the sky level by model fitting over
specified regions on either side of the spectrum.

The WHT data were wavelength calibrated using CuAr and CuNe arcs and
monitored using night sky lines. Arcs were taken at each object
position for wavelength calibrating the CTIO data.  We used solely the
sky lines to wavelength calibrate the Keck data. The spectra were flux
calibrated using observations taken of spectrophotometric standards.
B-stars free of strong features in the red were observed in order to
remove the effects of atmospheric absorption in the red-arm WHT
spectra and the CTIO spectra (e.g. O$_2$ A band at 7600 \AA). The
atmospheric absorption features seen in the B-star spectrum were
isolated by interpolating between values on either side of the
feature. The original B-star spectrum was then divided by this
atmospheric-free spectrum to create an atmospheric correction
spectrum. Finally the object spectra were divided by the scaled
correction spectrum. In the case of ISIS data the red and blue ends of
the spectra were then joined using SCOMBINE. In all cases if a
wide-slit observation was made (see Table~\ref{t_obs_log}), it was
used to correct the absolute flux levels for slit losses.

As mentioned in \S 2, the quasar spectra observed using the R-C
spectrograph at CTIO suffered a gradual flux decrement in the red end
calibration due to the inclusion of the second order flux from the
standard stars. In order to correct for this effect, spectra of
standard stars were taken with two different blocking filters (3450
\AA\ and 5000 \AA). The effect of the filter at wavelengths above 8000
\AA\ could thus be determined and a correction applied accordingly to
the quasar spectra. In addition a quasar previously observed with the
ISIS double-spectrograph on WHT was reobserved at CTIO and
corrected as explained above. Comparing the two spectra reveals no
significant difference and provides a successful double check on the
method. In any case the flux calibration of the red end of the spectra
is relatively unimportant for the work undertaken here, namely the
search of quasar absorbers blueward of the Ly-$\alpha$ emission.

The resulting spectra have a signal-to-noise ratio ranging from $\approx
10 -30$ at 7000 \AA. They are shown in Figure~\ref{f_spec} with
arbitrary flux scales. Ten of the quasars presented here exhibit
intrinsic broad absorption lines (BAL). The feature in all of the CTIO
spectra at 8900 \AA\space is due to bad columns. In the 1999 run, the
objects were moved along the slit between exposures so the effects of
the bad columns were spread over a slightly wider region of the
spectrum.

\section{Redshift and Magnitude Measurements}

In order to measure the redshifts, Gaussians were fit, if possible, to
N V (rest wavelength 1240.13 \AA), O I (1304.46 \AA), Si IV + O IV]
(1400.0 \AA) and C IV (1549.1 \AA) emission lines. The redshift for
each line was determined from the central wavelength ($z =
\lambda_{observed} / \lambda_{emitted} -1$). Ly$\alpha$ (rest
wavelength 1215.67 \AA) is almost 50 $\%$ absorbed by the Ly$\alpha$
forest, so that the blue edge of the emission line has been used for
redshift determination whenever possible. Some lines were impossible
to fit and made the redshift determination difficult, especially in
the case of the BALs. The redshift of each emission line, their
average and the 1 $\sigma$ error are shown in Table~\ref{t_z}. This
error is representative of the error in the fit, in wavelength
calibration (estimated to be around 0.1 \AA) and in the fact that the
various species are coming from different physical regions of the
quasar. In practice, this latter effect is probably the dominant
source of differences in the emission line redshifts.

To provide an internal check on our redshift determinations we
produced a median composite quasar spectrum. Each non-BAL spectrum
with enough wavelength coverage was corrected to the rest-frame and
scaled such that the median of the flux in a region free from emission
lines (1420-1470 \AA) is unity. The spectra were then rebinned into
fixed 0.5 \AA\ bins (i.e. similar to the resolution in the observed
frame) and the median of the flux in each bin was calculated to
produce the spectrum in Figure~\ref{f_medspec}. Measuring the
wavelength of the emission lines of the median composite spectrum
provides an estimate of any systematic bias in the redshift
measurements which in our case proves to be about z=0.001 (see last
row of Table~\ref{t_z}) and allows any shift in quasar redshifts to be
checked by cross-correlation.

Table~\ref{t_mag} summarizes the photometric and spectroscopic
magnitudes for each object. The photometric magnitudes were measured
from the UKST and POSS1 plates scanned using the APM facility. The
spectroscopic magnitudes are derived from the spectra themselves using
the IRAF task CALCPHOT and the ``R59'' (R) or ``R63'' (OR) filter
curves for the quasars magnitudes from the UKST survey and the ``e''
filter curve for quasar magnitudes from POSS1 survey. These transmission
curves are shown on Figure~\ref{f_filters} overplotted on the z=4.172
BR J0529$-$3552 quasar spectrum. The error on the spectroscopic
measurements is estimated to be +/- 0.1 mag and the error on the APM
R magnitude is +/-0.25 mag.

\section{Lyman-limit Systems}

\subsection{Background}
Lyman-limit systems (LLS) are absorption systems with hydrogen column
densities $N_{HI} \geq 1.6 \times 10^{17} $ atoms cm$^{-2}$. They are
defined by a sharp break (due to absorption of photons capable of
ionizing HI) shortward of 912 \AA. At $z < 1$, LLS are probably
associated with galactic halos (Steidel, Dickinson \& Persson 1994).

Because the search for absorption systems in quasar spectra is not
biased towards luminous intervening objects, our data constitute a
complementary way to the more traditional emission observations to
probe galaxy evolution. The observed number density of LLSs per unit
redshift can, for example, be directly compared with models of
structure formation (i.e. Abel \& Mo 1998). ``Grey'' LLS provide some
of the best candidates for measurement of the primordial abundance of
deuterium (Molaro \etal\ 1999, Levshakov \etal\ 2000 and references
herein). Subsequent papers will make use of this new sample of LLS to
constrain their space density, especially at $z\geq 2.4$ where we now
have significantly more data than in previous surveys (\eg\ Sargent,
Steidel, \& Boksenberg 1989; Lanzetta \etal\ 1991; Storrie-Lombardi
\etal\ 1994; Stengler-Larrea \etal\ 1995).

\subsection{LLS detection}

The LLS were identified using an automated technique which determines
the ratio $(f_+ / f_-)$ of the median of fluxes over 50 \AA\space
(rest frame) wide bins slid along the spectrum (see Schneider \etal\
1993 for details on this method). The minimum ratio corresponds to a
potential LLS detection and the redshift is calculated from the
corresponding wavelength. Two examples of the ratio versus LLS
redshift plots are shown in Figure~\ref{f_lls}, one for a detection
and the other for a non-detection. The optical depth is expressed as
the logarithm of the ratio previously defined: $\tau_{LLS} = - ln (f_+
/ f_-)$. In some cases, only a fraction of the radiation is absorbed
forming a step in the quasar spectra which does not reach zero flux
level. These so-called ``grey'' systems are only taken into account if
they have an optical depth, $\tau$, $> 1$. The redshifts and optical
depths of the LLS detected in our sample of quasars are summarized in
Table~\ref{t_lls}, together with the minimum and maximum redshift over
which a LLS {\it could} have been detected. The minimum redshift
corresponds to the smallest wavelength in the spectrum and the maximum
redshift is 3000 km s$^{-1}$ blueward of the quasar emission
redshift. The actual redshift path surveyed is usually limited by the
detection of the first Lyman-limit absorber, blueward of which there
is either no residual flux or insufficient signal-to-noise to detect
further LLSs. The analysis results in the detection of 49 LLS, 15 of
which are within 3000 km s$^{-1}$ of the quasar emission and thus
might be associated with the quasar itself. In some cases, metal
absorption features are also observed at the redshift of the LLS.

\section{Damped Lyman-alpha Candidates}

\subsection{Background}

Damped Lyman-alpha systems have, by definition (Wolfe \etal\ 1986), a
rest-frame equivalent width $W \geq 10$ \AA\ corresponding to
$N_{HI} \geq 2 \times 10^{20}$ atoms cm$^{-2}$.  At low redshift such
high HI column densities are found predominantly in gas rich systems
such as the disks of spiral galaxies. Kinematic studies (such as
Prochaska \& Wolfe 1998 and Ledoux \etal\ 1998) and metallicity
analyses (such as Pettini \etal\ 1997 and Prochaska \& Wolfe 2000)
indicate that DLAs at high-redshift might be the progenitors of
present day galaxies. Detecting these systems beyond $z \geq 4$
provides observational information about the early stages of galaxy
evolution.

DLAs are rare and to find them requires probing many quasar lines of
sight. Figure~\ref{f_g(z)} shows the cumulative number of lines of
sight along which a DLA {\it could} have been detected at the
5$\sigma$ confidence level.  This survey sensitivity, $g(z)$, is
compared with those of previous DLA surveys to show that our new
observations more than double the redshift path searched for DLAs at z
\simgt 3.5. Although DLAs have a low number density per unit redshift
compared with lower column density systems, they are thought to
contain most of the neutral hydrogen mass at z $ < $ 3. In a
subsequent paper (Peroux \etal\, in preparation), we will discuss how
this new survey impacts upon measurement of the comoving mass density
of neutral gas at high redshift, its implications for the formation
epoch of DLA and for the rate of evolution of gas into star.

\subsection{DLA Detection}

To select DLA candidates we have used the detection algorithm
following the method developed by Lanzetta \etal\ (1991), supplemented
by a visual search.  This has previously been applied to other samples
of z $>$ 4 quasars in Storrie-Lombardi \etal\ (1996a) and
Storrie-Lombardi \& Wolfe (2000).  We used the same method for fitting
the quasar continua as described in those papers. The spectra were
analyzed from 3000 km s$^{-1}$ blueward of the emission line (to avoid
lines possibly associated with the quasar) towards shorter
wavelengths. The analysis was stopped when the signal-to-noise ratio
became too low to detect a Ly$\alpha$ line with rest equivalent width
$\geq 5$ \AA\ at the 5$\sigma$ level (corresponding to $z_{min}$ in
Table~\ref{t_dlacand}). This point was typically caused by the
incidence of a Lyman limit system. We measured the equivalent widths
of all the candidates with rest equivalent widths greater than 5 \AA\
and estimated their {\nhi} column densities from the linear part of
the curve of growth. Previous experience has shown that the column
density estimates derived using this method are in good agreement with
measurements done on higher resolution data (compare Storrie-Lombardi
\etal\ (1996c) and Storrie-Lombardi \& Wolfe (2000)). The results
are listed in Table~\ref{t_dlacand}. The candidates with rest
equivalent widths in the range 5 -- 10 \AA\ at z$\sim$4 are listed in
the table for completion although they are unlikely to be damped.
 
Figure~\ref{f_finddla} shows two examples (BR J0006$-$6208 and BR
J0307$-$4945) of the output of the algorithm we use to detect DLA
candidates. The highest-redshift (z=4.46) DLA system currently known
is detected in the spectrum of quasar BR~J0307$-$4945
(Figure~\ref{f_q0307dla}). It has many associated metal lines which
have been studied in detail with higher-resolution observations
undertaken with the UVES spectrograph (Dessauges-Zavadsky \etal\
2000).

\subsection{Metal Lines in the DLAs}
Absorption features redward of the Ly-$\alpha$ quasar emission line
were selected using an automated algorithm~\footnote{see the following
URL for more details: http://www.ast.cam.ac.uk/$\sim$rfc/rdgen.html}
developed by Bob Carswell. The code systematically detects lines with
equivalent width $W \geq 0.1$ \AA. Gaussians were fitted to the lines
in order to measure their redshifts and equivalent widths. Some of
these lines were identified as low-ionization states of metals in
association with DLA candidates. In some cases Ly-$\beta$ was observed
blueward of the DLA candidate. All the metal lines associated with DLA
candidates are listed in Table~\ref{t_dlacand}.

\section{Metal Systems}
The observed equivalent width and wavelength of every absorption line
detected redward of the quasar Lyman-$\alpha$ emission were measured
using the algorithm described in section \S 6.3 above. The features
which were not associated with a DLA or LLS were identified using the
line list in Table~\ref{t_ionlist}. Most of the detected MgII systems
also show associated {\fe2} absorption. This survey resulted in the
detection of 80 new CIV systems (3.0 \simlt z \simlt 4.5) and 48 new
MgII systems (1.3 \simlt z \simlt 2.2). The results are summarized in
Table~\ref{t_metals}.

\section{Notes on Individual Objects}

\begin{enumerate}

\item PSS J0003$+$2730 (z=4.240): This quasar has two weak {\lya}
absorbers at z$=$3.51 and 3.89. Neither has an estimated column
density greater than $10^{20.3}$ atoms cm$^{-2}$, but
metal lines associated with both absorbers have been detected.

\item BR J0006$-$6208 (z=4.455): This quasar has weak emission lines
but a rich absorption spectrum. There are four candidate damped
absorbers at z$=$2.97, 3.20, 3.78 and 4.14. The highest redshift is a
weak candidate but the other three all have high estimated column
densities.  All the candidates have at least one associated metal
absorption line.  In addition, there is a MgII absorption system at
z$=$1.958.

\item BR J0030$-$5129 (z=4.174): This quasar has one candidate damped
absorber at z$=$2.45 with three associated {\fe2} lines.

\item PSS J0034$+$1639 (z=4.293): This quasar has two damped {\lya}
candidate absorbers.  The first is at z$=$3.75 and the estimated
column density of log {\nhi} $=$ 20.2 falls just below the formal
definition of DLAs. Associated {\sil2} and {\civ} metal lines are
detected.  The second damped system is at z$=$4.26 which is within
3000 {\kms} of the emission redshift of the quasar (z$=$4.293) so it
will not be included as an intervening absorber in the statistical
samples used in determining the neutral gas mass.  However it is of
interest because this is the first damped system detected at a
redshift greater than 4 with a column density log {\nhi} $>$ 21.  We
estimate the column density for this system at log {\nhi} $=$ 21.1 and
detected associated metals lines of \sil2, \o1, \c2, \si4, \civ, and
{\fe2} in the range z$=$4.252-4.282.
 
\item SDSS J0035$+$0040 (z=4.747): We detect no damped {\lya}
candidates in this spectrum.  This is one of the lower signal-to-noise
spectra in our sample due to the faintness of the quasar (R=21.3) but
we would have been able to detect a DLA with a column density log
{\nhi} $\ge 20.3$ over the redshift range 3.309 $<$ z $<$ 4.690.

\item PSS J0052$+$2405 (z=1.90): This is a broad absorption lines
quasar at z$=$1.9.  We observed it because the coordinates were
originally in the list of PSS $z > 4$ quasars available at their www
site~\footnote{http://www.astro.caltech.edu/$\sim$george/z4.quasars}.
It has since been removed from that list.

\item Q J0054$-$2742 (z=4.464): This quasar exhibits broad absorption
lines. The spectrum is not used in our absorption line survey.

\item PSS J0106$+$2601 (z=4.309): This quasar has a strong candidate
damped absorber at z$=$3.96 with associated metal lines.

\item PSS J0131$+$0633 (z=4.417): This quasar has two very weak
candidate damped absorbers at z$=$3.17 and 3.69.  {\civ} is also
detected at z$=$3.69.

\item PSS J0133$+$0400 (z=4.154): This spectrum has four candidate
damped absorbers.  The absorbers at z$=$3.69 and 3.77 have estimated
column densities above the formal definition of DLA ($N_{HI} \geq
10^{20.3}$ atoms cm$^{-2}$) and the absorbers at z$=$3.08 and 4.00 are
below that threshold.  Associated metal lines are detected for all of
the candidate DLAs.

\item PSS J0134$+$3307 (z=4.532): The quasar has a DLA at z=3.76 with
associated metal lines.

\item PSS J0137$+$2837 (z=4.258): This quasar exhibits broad
absorption lines. The spectrum is not used in our absorption line
survey.

\item PSS J0152$+$0735 (z=4.051): This quasar has an excellent DLA
candidate at z$=$3.84 with associated metal lines, which is also
detected as a Lyman limit system.
    
\item PSS J0207$+$0940 (z=4.136): This quasar exhibits strong
intrinsic absorption features. The spectrum is not used in our
absorption line survey.
   
\item PSS J0209$+$0517 (z=4.174): This quasar has weak emission lines
but exhibits two DLA candidates at z$=$3.66 and 3.86.  Both have
associated metal absorption features.
   
\item SDSS J0211$-$0009 (z=4.874): We detect one weak candidate DLA in
this quasar at z$=$4.64. {\sil2} is also detected at that redshift.
   
\item BR J0234$-$1806 (z=4.301): This quasar shows one weak absorption
candidate at z$=$3.69 with associated metal lines.
     
\item PSS J0248$+$1802 (z=4.422): This spectrum shows no DLA
candidates.
   
\item BR J0301$-$5537 (z=4.133): This quasar shows three DLA
candidates at z$=$3.22, 3.38, and 3.71.  All have associated metal
lines but the two higher redshift candidates have estimated column
densities below $2 \times 10^{20}$ atoms cm$^{-2}$.
   
\item BR J0307$-$4945 (z=4.728): The spectrum shows the highest
redshift damped absorber currently known at z$=$4.46 with an estimated
column density of log {\nhi} $=$ 20.8.  Associated metal lines of
\sil2, \o1, \c2, \si4, \civ, \fe2, and {\al2} are also detected at
this redshift.  The spectrum is shown in Figure~\ref{f_q0307dla}.  The
spectrum of this object is discussed in more detail in McMahon \etal,
in preparation and Dessauges-Zavadsky \etal\ (2000).  There is a weak
DLA candidate at z$=$3.35 with no associated metal lines.  This
absorber is highly unlikely to be damped.
   
\item SDSS J0310$-$0014 (z=4.658): This quasar shows two candidate
DLAs at z$=$3.42 and 4.34.  The lower redshift system has an estimated
column density above the DLA threshold. An associated
{\al2} line is detected at z$=$3.424 but no metal lines associated
with the higher redshift candidate are detected .
 
\item BR J0311$-$1722 (z=4.039): This quasar has a weak DLA candidate
at z$=$3.73 which is also detected as a Lyman-limit system. Associated
metal lines are also detected.

\item PSS J0320$+$0208 (z=3.840): This quasar exhibits broad
absorption lines. The spectrum is not used in our absorption line
survey.

\item BR J0324$-$2918 (z=4.622): No DLA candidates are detected in
this spectrum.

\item BR J0334$-$1612 (z=4.363): A DLA candidate at z$=$3.56 with
associated {\sil2} is detected in this quasar. This candidate has
previously been detected (Storrie-Lombardi \& Wolfe 2000) with a lower
estimated column density (log \nhi=20.6) than we measure .

\item SDSS J0338$+$0021 (z=5.010): This quasar has one DLA candidate
at z$=$4.06 with associated metals detected.

\item BR J0355$-$3811 (z=4.545): No DLA candidates are detected in
this spectrum.  There is a {\mgii} absorber at z$=$2.228.

\item BR J0403$-$1703 (z=4.227): No DLA candidates are detected. No
metal lines could be identified in this spectrum.

\item BR J0415$-$4357 (z=4.070): A weak DLA candidate with associated
metal lines is detected at z$=$3.81.

\item BR J0419$-$5716 (z=4.461): Three weak DLA candidates are
detected just above the Lyman-limit edge in this spectrum at z$=$2.82,
2.90, and 2.98. One associated metal line is detected from each of the
two lower redshift systems.

\item BR J0426$-$2202 (z=4.320): A very high column density candidate
(log \nhi=21.1) is detected at z$=$2.98 with associated {\al2}.

\item PMN J0525$-$3343 (z=4.383): No DLA candidates are detected in
this spectrum. Two {\mgii} absorbers are detected at z$=$1.570 and
2.006.

\item BR J0529$-$3526 (z=4.413): A weak DLA candidate with associated
metal lines is detected at z$=$3.57.

\item BR J0529$-$3552 (z=4.172): A `doublet' of weak DLA candidates is
detected at z$=$3.68 and 3.70. No associated metals are detected at
these redshifts.
     
\item BR J0714$-$6455 (z=4.462): No DLA candidates are detected in
this spectrum.

\item PSS J0747$+$4434 (z=4.430): Two DLA candidates are detected at
z$=$3.76 and 4.02. The higher redshift system also has associated
metal lines.

\item RX J1028$-$0844 (z=4.276): Two weak DLA candidates with
associated metals are detected at z$=$3.42 and 4.05.

\item PSS J1048$+$4407 (z=4.381): This quasar exhibits broad
absorption lines. The spectrum is not used in our absorption line
survey.

\item PSS J1057$+$4555 (z=4.116): Three DLA candidates are detected at
z$=$2.90, 3.05, and 3.32.  The candidate absorber at z$=$3.32 has been
confirmed as damped in a higher resolution spectrum. It has a redshift
of z$=$3.3172 and a column density log \nhi=20.34 (Lu, Sargent \&
Barlow 1998).  The estimated column density (log \nhi=20.3) for the
z$=$3.05 is identical to the estimate reported in Storrie-Lombardi \&
Wolfe (2000).

\item PSS J1159$+$1337 (z=4.073): This quasar has a DLA candidate at
z$=$3.72 with several associated metal lines.
 
\item PSS J1253$-$0228 (z=4.007): This quasar has two candidate damped
absorbers.  One absorber at z$=$2.78 has a very high estimated column
density (log \nhi=21.4) and an associated {\al2} line is
detected. This is the highest column density absorber in our
survey. Another absorber at z$=$3.60 is highly unlikely to be damped,
with an estimated column density of log \nhi=19.7, but does have
several associated metal lines.

\item BR J1310$-$1740 (z=4.185): This quasar has a weak damped
candidate at z$=$3.43.  Associated metal lines are also detected at
this redshift.

\item BR J1330$-$2522 (z=3.949): This quasar has two weak DLA
candidates at z$=$2.91 and 3.08.  The higher redshift system has
associated metal lines.

\item FIRST J1410$+$3409 (z=4.351): There is a weak candidate damped
absorber at z$=$3.43 with no associated metal lines. In this spectrum
the redshift path surveyed for damped absorbers is not continuous due
to a large noise spike in the forest at z$\approx$3.59.

\item PSS J1438$+$2538 (z=4.234): This quasar exhibits broad
absorption lines.  The spectrum is not used in our absorption line
survey.

\item PSS J1456$+$2007 (z=4.249): There are two weak DLA candidates at
z$=$ 3.22 and 4.16.  The lower redshift system also has associated
metal lines.

\item BR J1603$+$0721 (z=4.385): No DLA candidates are detected in
this spectrum.

\item PSS J1618$+$4125 (z=4.213): There is a DLA candidate at z$=$3.92
with associated metal lines.

\item PSS J1633$+$1411 (z=4.351): There is a weak DLA candidate at
z$=$3.90 with associated metal lines.

\item PSS J1646$+$5514 (z=4.037): No DLA candidates are detected in
this spectrum.

\item PSS J1721$+$3256 (z=4.031): No DLA candidates are detected in
this spectrum.

\item RX J1759$+$6638 (z=4.320): There is a DLA candidate at z$=$3.40
with associated metal lines.
    
\item PSS J1802$+$5616 (z=4.158): There are four damped absorber
candidates detected in this spectrum at z$=$3.39, 3.56, 3.76, and
3.80.  Only the absorber at z$=$3.76 has an estimated column density
above the formal definition of DLA ($N_{HI} \geq 10^{20.3}$ atoms
cm$^{-2}$).  Associated metal lines are detected for the z$=$3.39 and
3.80 absorbers.

\item BR J2017$-$4019 (z=4.131): This quasar exhibits strong intrinsic
absorption at the quasar emission redshift.  The {\civ} and {\si4}
emission lines are completely absorbed.  The spectrum is not used in
our absorption line survey.

\item PSS J2122$-$0014 (z=4.114): This spectrum shows two DLA
candidates at z$=$3.20 and 4.00. We estimate the column density of the
lower redshift system to be log {\nhi}=20.3, but this may be an
overestimating as the {\lya} line at z$=$3.20 is at the same position
as the {\lyb} line at z$=$4.00. Associated metal lines are detected
for both absorption systems.

\item BR J2131$-$4429 (z=3.834): This quasar exhibits broad absorption
lines. The spectrum is not used in our absorption line survey.

\item PMN J2134$-$0419 (z=4.334): This quasar has one weak DLA
candidate at z$=$3.27 with associated metal lines.

\item PSS J2154$+$0335 (z=4.363): This quasar has two DLA candidates
at z$=$3.61 and 3.79.  Metal lines are detected for both, but only the
lower redshift system has an estimated column density above $2 \times
10^{20}$ atoms cm$^{-2}$.

\item PSS J2155$+$1358 (z=4.256): This quasar has a very high column
density (log {\nhi} $=$ 21.1) DLA candidate at z$=$3.32.  Associated
metal lines are also detected at this redshift.

\item BR J2216$-$6714 (z=4.469): This quasar has three weak DLA
candidates at z$=$3.27, 4.28, and 4.32.  At least one associated metal
line has been detected for each.

\item PSS J2241$+$1352 (z=4.441): This quasar has two DLA candidates
at z$=$3.65 and 4.28.  The lower redshift system has an estimated
column density below the formal definition of DLA ($N_{HI} \geq
10^{20.3}$ atoms cm$^{-2}$). Both have associated metal lines.

\item DMS B2247$-$0209 (z=4.335): This quasar exhibits broad
absorption lines. The spectrum is not used in our absorption line
survey.

\item PSS J2315$+$0921 (z=4.412): This quasar exhibits strong
intrinsic absorption at the quasar emission redshift.  The {\civ} and
{\si4} emission lines are almost completely absorbed.  It is similar
in character to the spectrum of BR J2017$-$4019.  The spectrum is not
used in our absorption line survey.

\item BR J2317$-$4345 (z=3.943): This quasar has a strong DLA
candidate at z$=$3.49 with associated metal lines.

\item BR J2328$-$4513 (z=4.359): There is a weak DLA candidate at
z$=$3.04.  {\sil2} is detected at this redshift but may be blended
with {\civ} at z$=$3.719.

\item PSS J2344$+$0342 (z=4.239): There are two very high column
density DLA candidates at z$=$2.68 and 3.21.  Both have associated
metal lines.

\item BR J2349$-$3712 (z=4.208): There is a weak DLA candidate at
z$=$3.69 with associated {\sil2}.

\end{enumerate}

\section{Conclusions} 
We have presented the spectra ofsixty-six6 $z \ga 4$ bright quasars with
$\sim 5$ \AA\ resolution (FWHM) and signal-to-noise ratio ranging from
10 to 30. Twenty-six new damped Ly$\alpha$ absorption candidates
(column densities $N_{HI} \geq 2 \times 10^{20}$ atoms cm$^{-2}$) and
forty-nine new Lyman-limit systems ($N_{HI} \geq 1.6 \times 10^{17}$
atoms cm$^{-2}$), fiften of which are within 3000 km s$^{-1}$ of the
quasar emission, have been discovered. The space density and column
density evolution of these systems will be analyzed in subsequent
papers. Higher resolution observations are needed in order to confirm
the column density measurement and to differentiate the multiple
systems among the DLA candidates presented here, and in order to make
detailed analysis of the metallicity content of these absorbers. These
high-column density systems can also be used to measure the neutral
hydrogen content of the Universe over a large redshift range, thus
probing the formation epoch of these objects and tracing the gas from
which stars form. Analyzed in conjunction with previous studies, our
new survey will provide enough data to help draw statistically more
significant conclusions on these issues at high redshift.

\acknowledgements The authors are indebted to the referee, Julia
Kennefick, for helpful comments. CP warmly thanks Max Pettini for
useful suggestions on an earlier version of this manuscript, Bob
Carswell for help with his code and PPARC \& the Isaac Newton Trust
for support. LSL is grateful to the staff at CTIO for their expert
assistance in obtaining some of the observations presented here. RGM
would like to thank the Royal Society for support. This research has
made use of the NASA/IPAC Extragalactic Database (NED) which is
operated by the Jet Propulsion Laboratory, California Institute of
Technology, under contract with the National Aeronautics and Space
Administration. This paper is based on observations obtained at the
William Herschel Telescope which is operated on the island of La Palma
by the Isaac Newton Group in the Spanish Observatorio del Roque de los
Muchachos of the Instituto de Astrofisica de Canarias, on observations
made at the Cerro Tololo Intra-American Observatory which is operated
by the Association of Universities for Research in Astronomy, under a
cooperative agreement with the National Science Foundation as part of
the National Optical Astronomy Observatories and on data obtained at
the W.M. Keck Observatory, which is operated as a scientific
partnership among the California Institute of Technology, the
University of California and the National Aeronautics and Space
Administration.  The Observatory was made possible by the generous
financial support of the W.M. Keck Foundation.

\begin{deluxetable}{lccccc}
\tablewidth{0pc}
\tablenum{1}
\label{t_obs_log}
\tablecaption{Journal of Observations}
\tablehead{
\colhead{Quasar} &  \colhead{Telescope}&\colhead{Date} &\colhead{Exp. Time} &\colhead{Wide} &\colhead{Ref}  \\
\colhead{Name}& \multicolumn{2}{c}{Observed} &\colhead{B/R\tablenotemark{a} (secs)}&\colhead{Slit} & 
}
\startdata
PSS J0003$+$2730   &WHT  &1998 Sep 22 &3600/3600      &yes   &1\\	
BR J0006$-$6208    &CTIO  &1998 Oct 14  &3600  &yes          &3	\\	
BR J0030$-$5129    &CTIO  &1998 Oct 15  &3600  &yes          &3	\\	
PSS J0034$+$1639   &WHT  &1998 Sep 22 &3600/3600      &yes   &1\\	
SDSS J0035$+$0040  &CTIO  &1999 Oct 10 &8100 &yes &           4\\			
PSS J0052$+$2405   &WHT  &1998 Sep 23 &3600/3600      &yes   &2\\	
Q J0054$-$2742     &CTIO  &1999 Oct 12 &2700 &yes            &5 \\			
PSS J0106$+$2601   &WHT  &1998 Sep 24 &3600/3600      &yes   &1\\
\enddata
\tablenotetext{a}{For the quasars observed at the WHT the B/R designations give the exposure
times through the blue and red arms of the ISIS spectrograph.}
\tablecomments{The quasar prefixes indication the following origin:  BR = APM survey objects selected by B$_{\rm J}$-R 
color excess; PSS = Second Palomar Sky Survey; PMN = Parkes-MIT-NRAO radio-selected objects; RX = X-ray selected; 
SDSS  = Sloan Digital Sky Survey; and DMS  = Deep Multicolor Survey} 
\tablerefs{ 
(1) Stern et~al.~2000;
(2) G. Djorgovski's www page at http://www.astro.caltech.edu/$\sim$george/z4.qsos;
(3) Storrie-Lombardi et~al.~2000;
(4) Fan et~al.~1999;                           
(5) Warren, Hewett, \& Osmer 1991;
(6) Kennefick et~al.~1995a;                      
(7) Storrie-Lombardi et~al.~1996a; 
(8) Zickgraf et~al.~1997;
(9) Kennefick et~al.~1995b;                      
(10) Henry et~al.~1994;
(11) Hook et~al.~in preparation;
(12) Hall et~al.~1996
}
\end{deluxetable}

\dummytable{\label{t_obs_log}}


\begin{deluxetable}{lccccc|cl}
\tablewidth{0pc}
\tablenum{2}
\label{t_z}
\tablecaption{Quasar Redshift Measurements}
\tablehead{
\colhead{Quasar} &  \colhead{\lya}&\colhead{NV} &\colhead{OI }
&\colhead{Si+O} &\colhead{CIV} &\colhead{Mean}&\colhead{Note} \\
&\colhead{1216\AA} &\colhead{1240\AA}&\colhead{1304\AA} &\colhead{1400\AA} &\colhead{1549\AA}&\colhead{Redshift\tablenotemark{a}}&
}
\startdata
PSS J0003$+$2730  &4.255  &4.205  &4.261  &4.248  &4.233  &4.240$\pm$0.022&\\      
BR J0006$-$6208  &4.505  &\nd &4.471  &\nd &4.388  &4.455  $\pm$.060&\\      
BR J0030$-$5129  &4.163  &4.172  &4.187  &4.175  &4.174  &4.174  $\pm$.009&\\      
PSS J0034$+$1639  &4.309  &\nd &4.298  &4.281  &4.284  &4.293  $\pm$.013&\\      
SDSS J0035$+$0040  &4.737  &\nd &4.758  &\nd &4.746  &4.747  $\pm$.011&\\      
PSS J0052$+$2405  &\nd &\nd &\nd &\nd &\nd &1.90  $\pm$0.05 & BAL, \tablenotemark{b}\\   
Q J0054$-$2742  &4.464  & \nd& \nd& \nd&\nd &4.464  $\pm$.005  &BAL  \\ 
PSS J0106$+$2601  &4.323  &\nd &4.332  &4.303  &4.276  &4.309  $\pm$.025&\\
\enddata 
\tablenotetext{a}{The error estimate is
$\sigma \over \sqrt n$.}  
\tablenotetext{b}{This object
was originally in the PSS web page as a high redshift object.  It has
since been removed from that list.  Our spectrum shows that it is a
broad absorption line quasar at z$=$1.9.}  
\tablecomments{BAL: These quasars exhibit
broad absorption line characteristics.}  
\end{deluxetable}

\dummytable{\label{t_z}}

\begin{deluxetable}{lcclc}
\tablewidth{0pc}
\tablenum{3}
\label{t_mag}
\tablecaption{Quasar Magnitude Measurements}
\tablehead{
\colhead{Quasars} &  \colhead{R\tablenotemark{a} }&\colhead{R\tablenotemark{b}} 
&\colhead{Filter\tablenotemark{c}} &\colhead{Notes} \\
 & APM  & &  
}
\startdata
PSS J0003$+$2730   &18.3	  &17.8  	&e    	&\\      
BR J0006$-$6208    &18.3	  &19.2    & R59   &\\      
BR J0030$-$5129    &18.6	  &18.6    & R59   &\\      
PSS J0034$+$1639   &18.0	  &17.8  	&e    	&\\      
SDSS J0035$+$0040  &\nd	  &21.3  	&e    	&\\      
PSS J0052$+$2405   &17.4	  &17.4  	&e    	&BAL\\       
Q J0054$-$2742     &19.8	  &19.8    & R59   &BAL\\       
PSS J0106$+$2601   &18.5	  &18.3  	&e    	&\\
\enddata
\tablenotetext{a} {The R magnitude from the photographic plates
as measured by the APM. If no magnitudes are specified in this 
column, the quasar is not detected on the APM plates. The
uncertainties in these magnitudes are estimated to be $\pm 0.25$.}
\tablenotetext{b} {The R magnitude measured from the spectra as 
described in the text. The
uncertainties in these magnitudes are estimated to be $\pm 0.1$.} 
\tablenotetext{c} {R59 and R63 are UKST filters while e is the POSS1 filter.}  
\tablecomments{The NC in the notes column means that  
the spectrum has not been corrected for slit losses. The BAL designation means
the quasar exhibits broad absorption lines.}  
\end{deluxetable}

\dummytable{\label{t_mag}}

\begin{deluxetable}{lccccc}
\tablewidth{0pc}
\tablenum{4}
\label{t_lls}
\tablecaption{Survey for Lyman-limit Systems}
\tablehead{
\colhead{Quasar} &  \colhead{z$_{em}$}&\colhead{z$_{min}$\tablenotemark{a}}
&\colhead{z$_{max}$\tablenotemark{b}} &\colhead{z$_{lls}$} &\colhead{$\tau$} 
}
\startdata
PSS J0003$+$2730	 &4.240		 &2.858		 &4.198		 &3.97	 &2.6\\
BR J0006$-$6208	 &4.455		 &3.079		 &4.400		 &4.14	 &1.8\\
BR J0030$-$5129	 &4.174		 &3.079		 &4.122		 &3.37	 &1.0\\
PSS J0034$+$1639	 &4.293		 &2.858		 &4.240		 &4.26  & 4.1\\
SDSS J0035$+$0040\tablenotemark{c}	 &4.747		 &3.079		 &4.690		 &4.59	 &0.9\\
PSS J0106$+$2601\tablenotemark{d}	 &4.309		 &2.858		 &4.256		 &4.05	 &2.3\\
		 &		 &		 &		 &3.96	 &2.7\\
PSS J0131$+$0633	 &4.417		 &3.079		 &4.363		 &4.37  & 2.1\\
\enddata	 
\tablenotetext{a}{$z_{min}$
corresponds to the smallest wavelength in the quasar spectrum.}
\tablenotetext{b}{$z_{max}$ is 3000 km s$^{-1}$ blueward of the quasar emission
redshift.}
\tablenotetext{c}{Systems with optical depth, $\tau$, $< 1$ are excluded from the total count of LLS because they do not conform to the formal definition of Lymn-limit Systems.}
\tablenotetext{d}{In case where two breaks were detected, only the highest system is taken into account in the final count for LLS.}
\tablenotetext{e}{No LLS have been detected over the specified redshift range.}
\tablenotetext{f}{The signal-to-noise ratio of this spectrum is too low to unable LLS detection.}
\end{deluxetable}

\dummytable{\label{t_lls}}

\begin{deluxetable}{lccccrclcl}
\tablewidth{0pc}
\tablenum{5}
\label{t_dlacand}
\tablecaption{Survey for Damped Lyman$-\alpha$ Absorption Systems}
\tablehead{
\colhead{Quasar} & \colhead{z$_{em}$}& \colhead{z$_{min}$}&\colhead{z$_{max}$}&\colhead{z$_{abs}$}& 
\colhead{W} &  \colhead{log \nhi} & \colhead{Metal} & \colhead{z$_{\rm metal}$} & \colhead{Note}\\
 & & & & & \colhead{(\AA)}& \colhead{(cm$^{-2}$)} & \colhead{Lines} &  &
}
\startdata
PSS  J0003$+$2730 & 4.240 & 2.718 & 4.188  & 3.51 &  7.6 & 20.0 & {\sil2} 1527   & 3.513 &   \\
                  &       &       &        &      &      &      & {\fe2}  1608   & 3.510 &   \\
                  &       &       &        &      &      &      & {\al2}  1671   & 3.512 &   \\
                  &       &       &        & 3.89 &  9.0 & 20.2 & {\c2}   1334   & 3.893 &   \\
                  &       &       &        &      &      &      & {\si4}  1400   & 3.893 &   \\
                  &       &       &        &      &      &      & {\civ}  1549   & 3.893 &   \\
                  &       &       &        &      &      &      & {\al2}  1671   & 3.891 &   \\
BR  J0006$-$6208  & 4.455 & 2.944 & 4.400  & 2.97 & 15.6 & 20.7 & {\sil2} 1808   & 2.965 &  \\
\enddata
\tablenotetext{a}{{\fe2} 1608 at z$=$3.780 is at the same position as {\fe2} 2600 at z$=$1.958.}
\tablenotetext{b}{Also detected as a Lyman-limit system.} 
\tablenotetext{c}{This damped system is within 3000 km s$^{-1}$ of the quasar emission 
redshift but we have included it in this table due to the fact that it is the
first damped absorber detected at a redshift z $>$ 4 with a column density log {\nhi} $>$ 21.}
\tablenotetext{d}{{\sil2} 1527 at z$=$3.761 is at the same position as {\civ} 1549 at z$=$3.686.}
\tablenotetext{e}{{\al2}  1671 at z$=$4.047 is at the same position as {\mg1} 2853 at z$=$1.956.}
\tablenotetext{f}{This damped absorption line has a very narrow core but strong damping wings
are visible on both sides of the line.}
\tablenotetext{g}{The {\lya} line may be blended with {\lyb} at z$=$4.00, therefore the column
density may be overestimated.  This quasar has an unusually rich absorption spectrum, with 
many {\civ} absorbers redward of the {\lya} emission.} 
\tablenotetext{h}{{\sil2} 1527 at z$=$3.316 is at the same position as {\fe2} 2260 at z$=$1.915.} 
\tablenotetext{i}{{\sil2} 1808 at z$=$3.041 is blended with {\civ} 1549 at z$=$3.719.} 
\tablenotetext{j}{This damped absorption candidate is just below the minimum redshift 
determined with our detection algorithm. 
It is likely to be real but requires confirmation with a higher signal-to-noise spectrum.}
\end{deluxetable}

\dummytable{\label{t_dlacand}}


\begin{deluxetable}{llc}
\tablewidth{0pc}
\tablenum{6}
\label{t_ionlist}
\tablecaption{Metal Lines Rest Wavelengths} 
\tablehead{
\colhead{Ion} &  \colhead{$\lambda$ (\AA)} & \colhead{Ref.}
}
\startdata
\n5     & 1238.821	&1\\       
\n5     & 1242.804	&1\\       
\sil2    & 1260.4221    &1\\     
\sil2    &1304.3702	&1\\     
\c2     &1334.5323      &1\\
\si4    &1393.7550      &1\\ 
\si4    &1402.7700      &1\\ 
\sil2    &1526.7066     &1\\
\civ     &1548.1950     &1\\ 
\civ     &1550.7700     &1\\ 
\fe2    &1608.4511      &1\\
\fe2    &1611.2005      &1\\
\al2    &1670.7874      &1\\
\sil2    &1808.0126     &2\\
\fe2    &2260.7805      &1\\
\fe2    &2344.2140      &1\\ 
\fe2    &2367.5910      &3\\    
\fe2    &2374.4612      &4\\
\fe2    &2382.7650      &4\\ 
\fe2    &2586.6500      &4\\
\fe2    &2600.1729      &4\\
\mgii    &2796.3520     &1\\  
\mgii    &2803.5310     &1\\  
\mg1    &2852.9641      &1\\   
\enddata
\tablerefs{
(1) Morton 1991;
(2) Bergeson \& Lawler 1993;
(3) Nussbaumer, Pettini \& Storey 1982;
(4) Cardelli \& Savage 1995
}
\end{deluxetable}

\dummytable{\label{t_ionlist}}


\begin{deluxetable}{lccclcc}
\tablewidth{0pc}
\tablenum{7}
\label{t_metals}
\tablecaption{Identification of Metal Absorption Lines}
\tablehead{ 
\colhead{Quasar} & \colhead{z$_{em}$} & \colhead{$\lambda_{\rm obs}$}& 
\colhead{W$_{\rm obs}$} & \colhead{Ion} &  \colhead{$\lambda_{\rm rest}$} &  \colhead{z} \\
 & & \colhead{(\AA)} & \colhead{(\AA)} & &  \colhead{(\AA)} & 
}
\startdata
PSS J0003$+$2730   & 4.240 & 6530.3 & 5.6  & \c2   & 1334  & 3.893\\ 
                   &       & 6819.2 & 2.1  & \si4  & 1393  & 3.893\\ 
                   &       & 6863.6 & 1.6  & \si4  & 1402  & 3.893\\ 
                   &       & 6889.8 & 1.4  & \sil2 & 1527  & 3.513\\ 
                   &       & 7254.7 & 1.7  & \fe2  & 1608  & 3.510\\ 
                   &       & 7320.6 & 1.3  & \civ  & 1548  & 3.729\\ 
                   &       & 7332.6 & 0.8  & \civ  & 1551  & 3.728\\ 
                   &       & 7468.7 & 3.3  & \sil2 & 1527  & 3.892\\
\enddata
\end{deluxetable}

\dummytable{\label{t_metals}}



\epsscale{.75}
\plotone{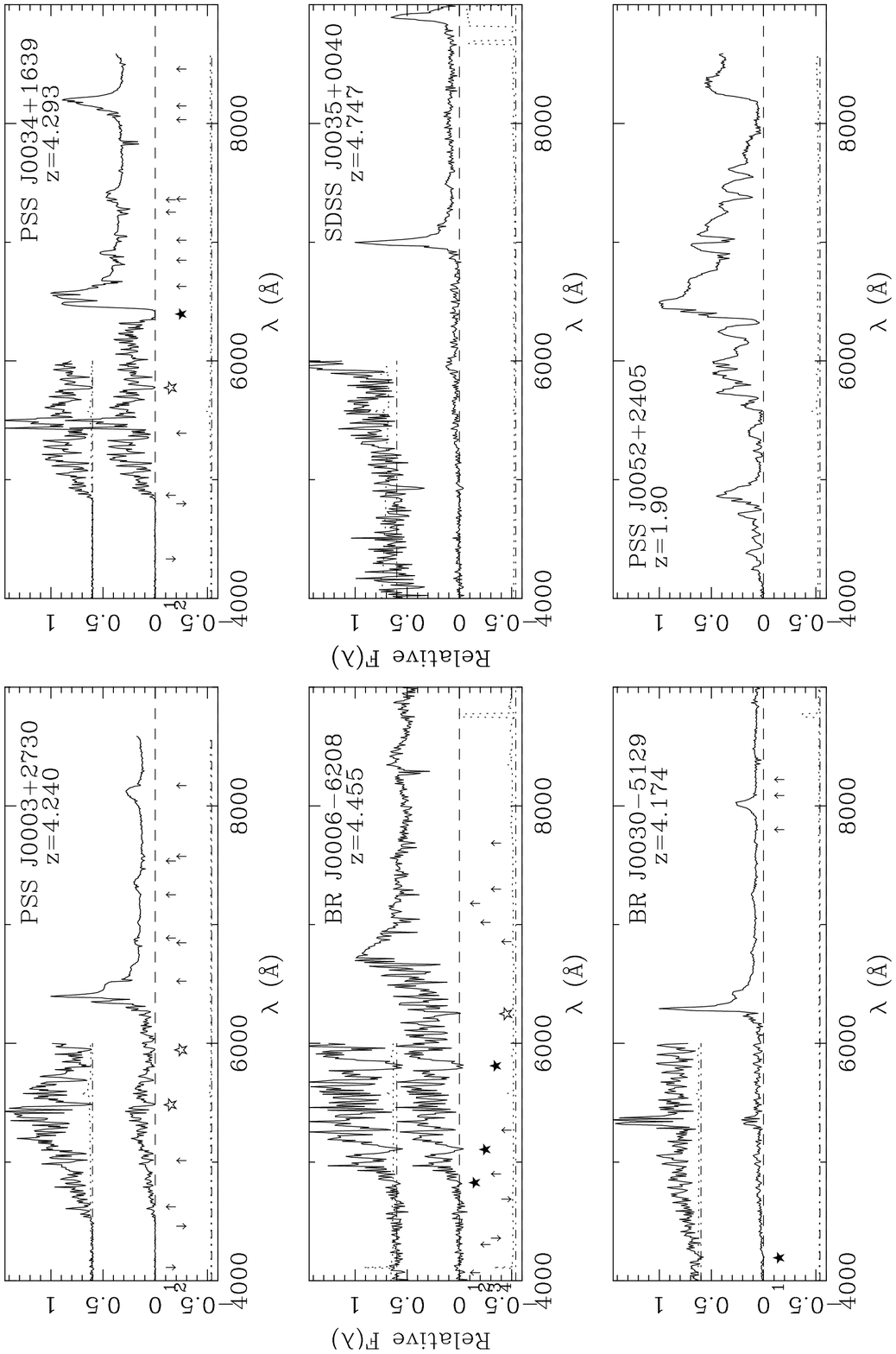}
\figcaption{Spectra of all the observed quasars.  The error arrays are
plotted as dotted lines, offset below the spectra for clarity.  In the
upper left-hand corner, the blue region of the spectra are magnified
to make the Lyman-limit systems and damped {\lya} candidate absorbers
easier to see.  Damped {\lya} candidate absorbers are marked below
there positions as solid stars if they have estimated column densities
{\nhi} $\ge 2 \times 10^{20}$ {\atomcm}, and as open stars if they
have estimated column densities lower than this threshold, but greater
than $5 \times 10^{19}$ {\atomcm}. To the right of the stars marking
the DLA are the detected metal lines that are associated with this
absorber.  To the left of the stars are an upward arrow marking the
position of {\lyb} at the DLA redshift and a downward arrow marking
the wavelength of the Lyman-limit that would be associated with this
DLA.
\label{f_spec}}
                    
\epsscale{.8}
\plotone{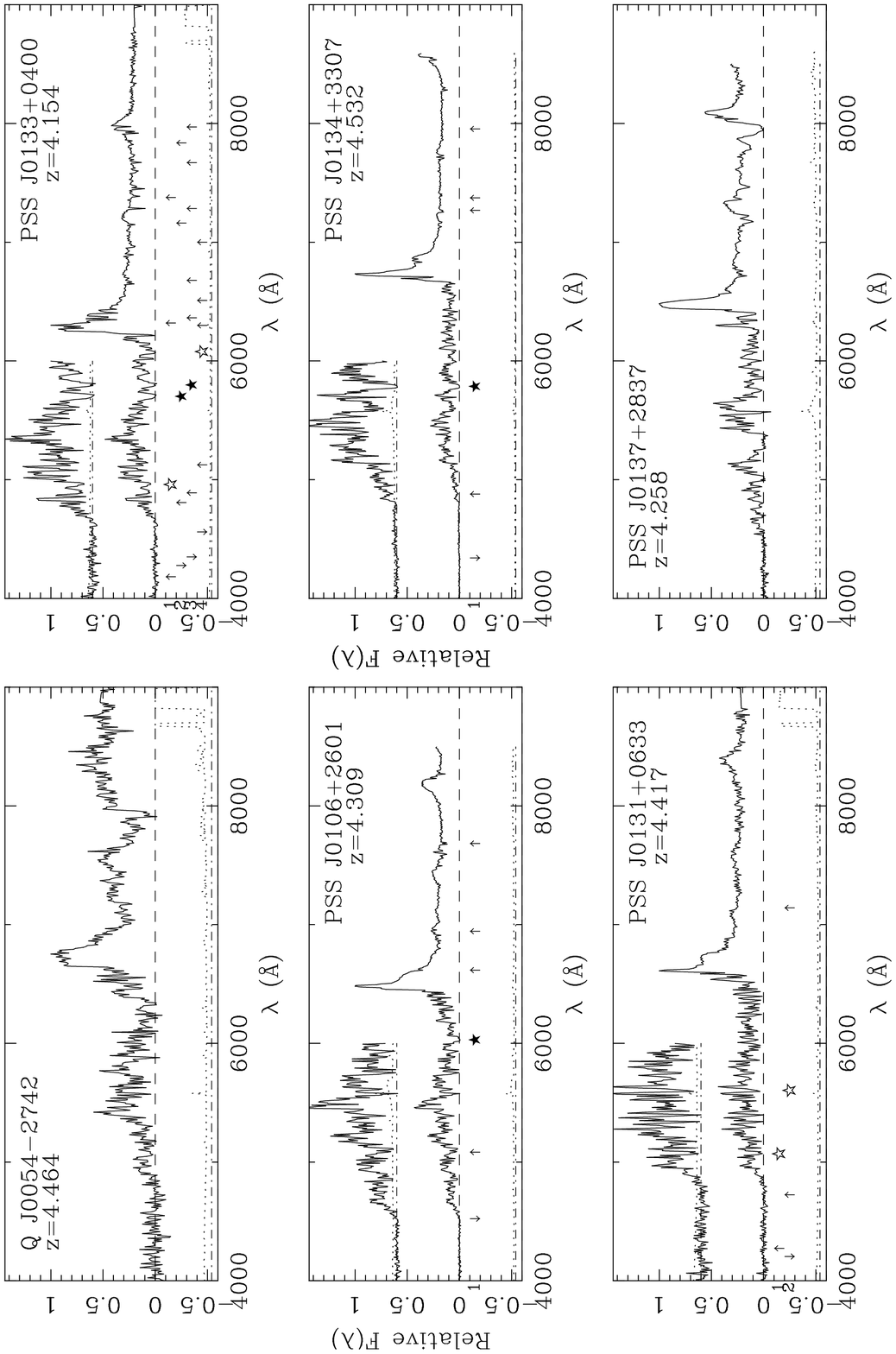}
\setcounter{figure}{0}
\figcaption{{\it continued}}
                     
\epsscale{.8}
\plotone{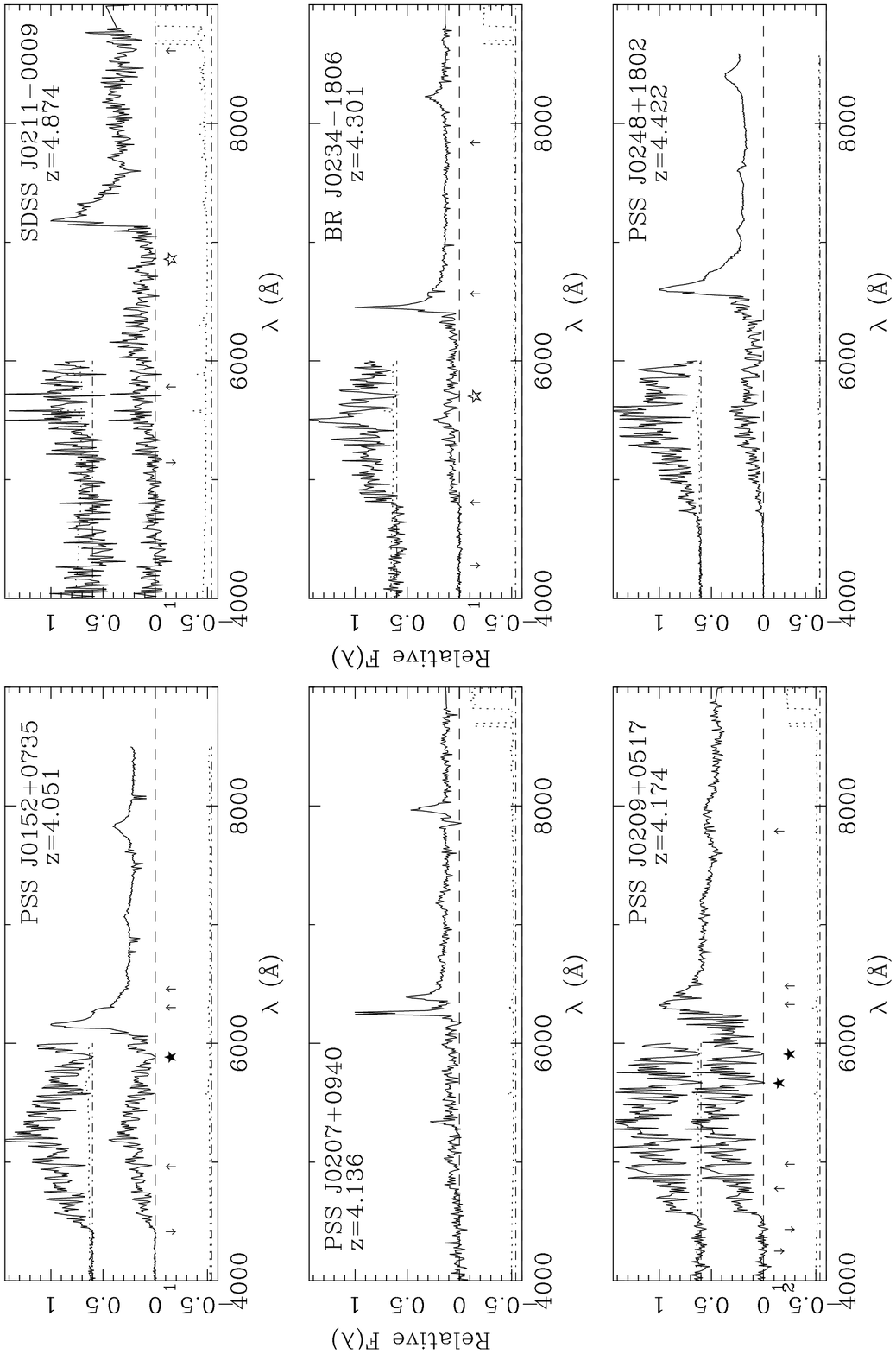}
\setcounter{figure}{0}
\figcaption{{\it continued}}
                     
\epsscale{.8}
\plotone{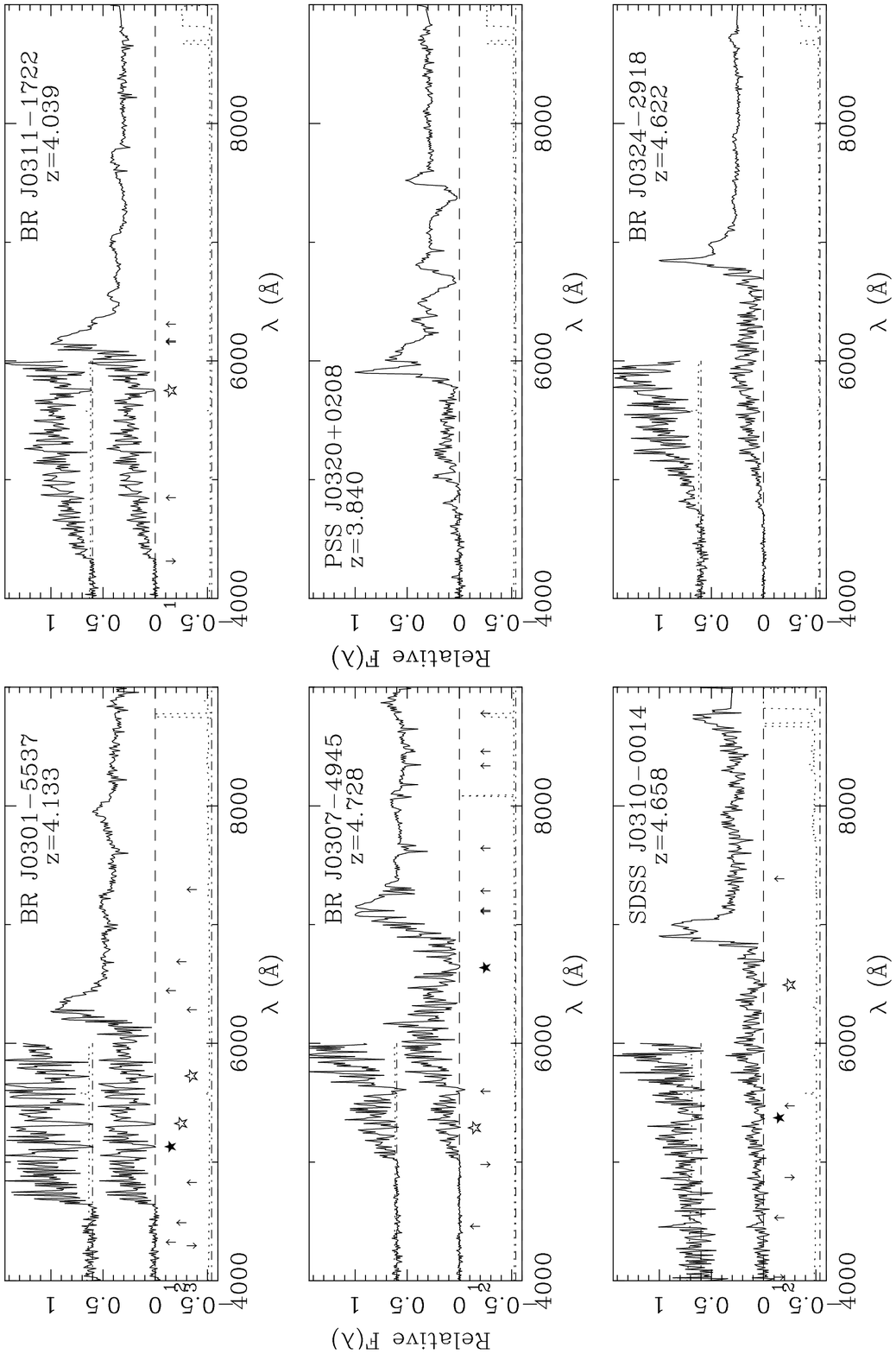}
\setcounter{figure}{0}
\figcaption{{\it continued}}
                     
\epsscale{.8}
\plotone{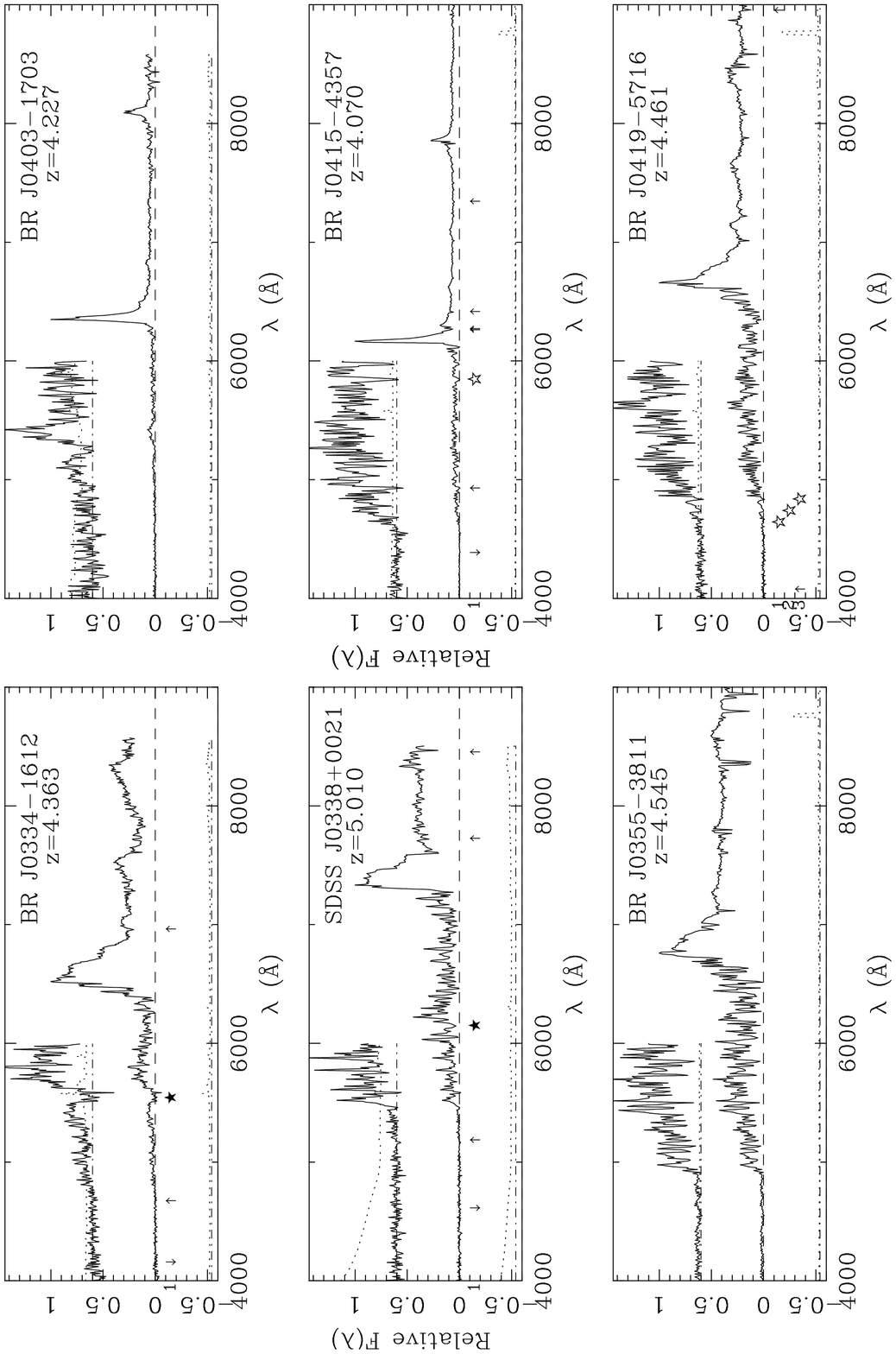}
\setcounter{figure}{0}
\figcaption{{\it continued}}
                     
\epsscale{.8}
\plotone{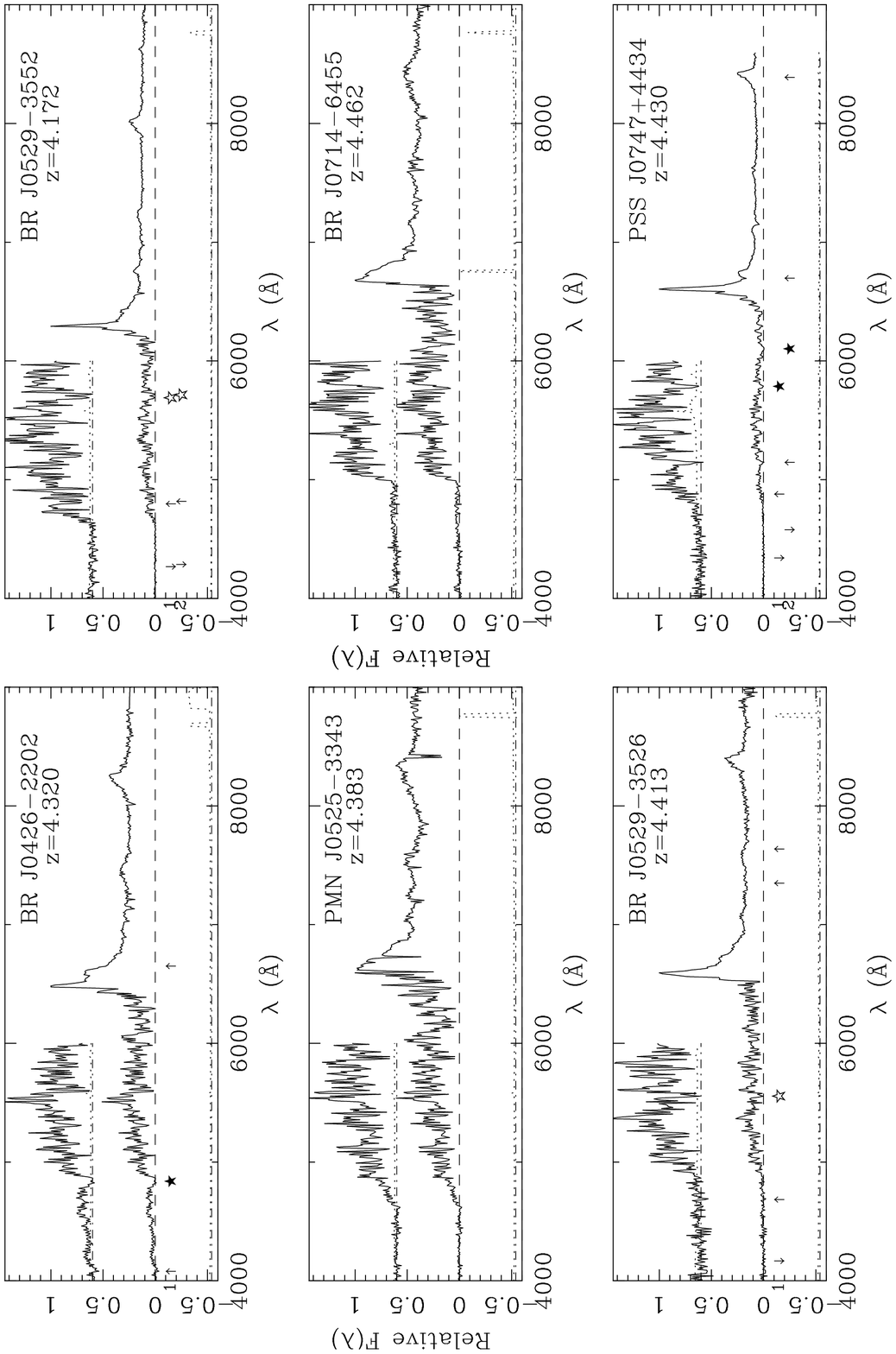}
\setcounter{figure}{0}
\figcaption{{\it continued}}
                     
\epsscale{.8}
\plotone{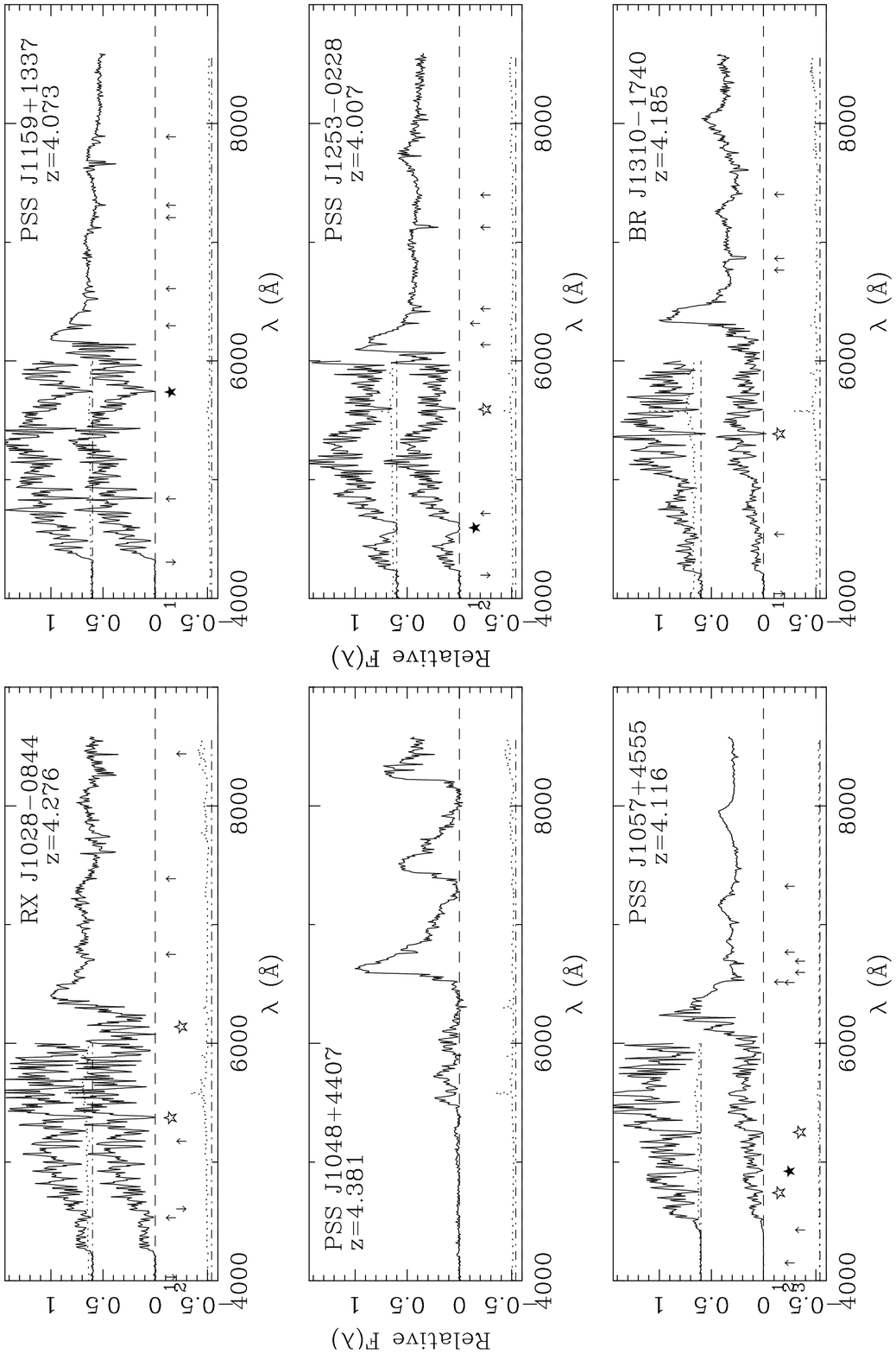}
\setcounter{figure}{0}
\figcaption{{\it continued}}
                     
\epsscale{.8}
\plotone{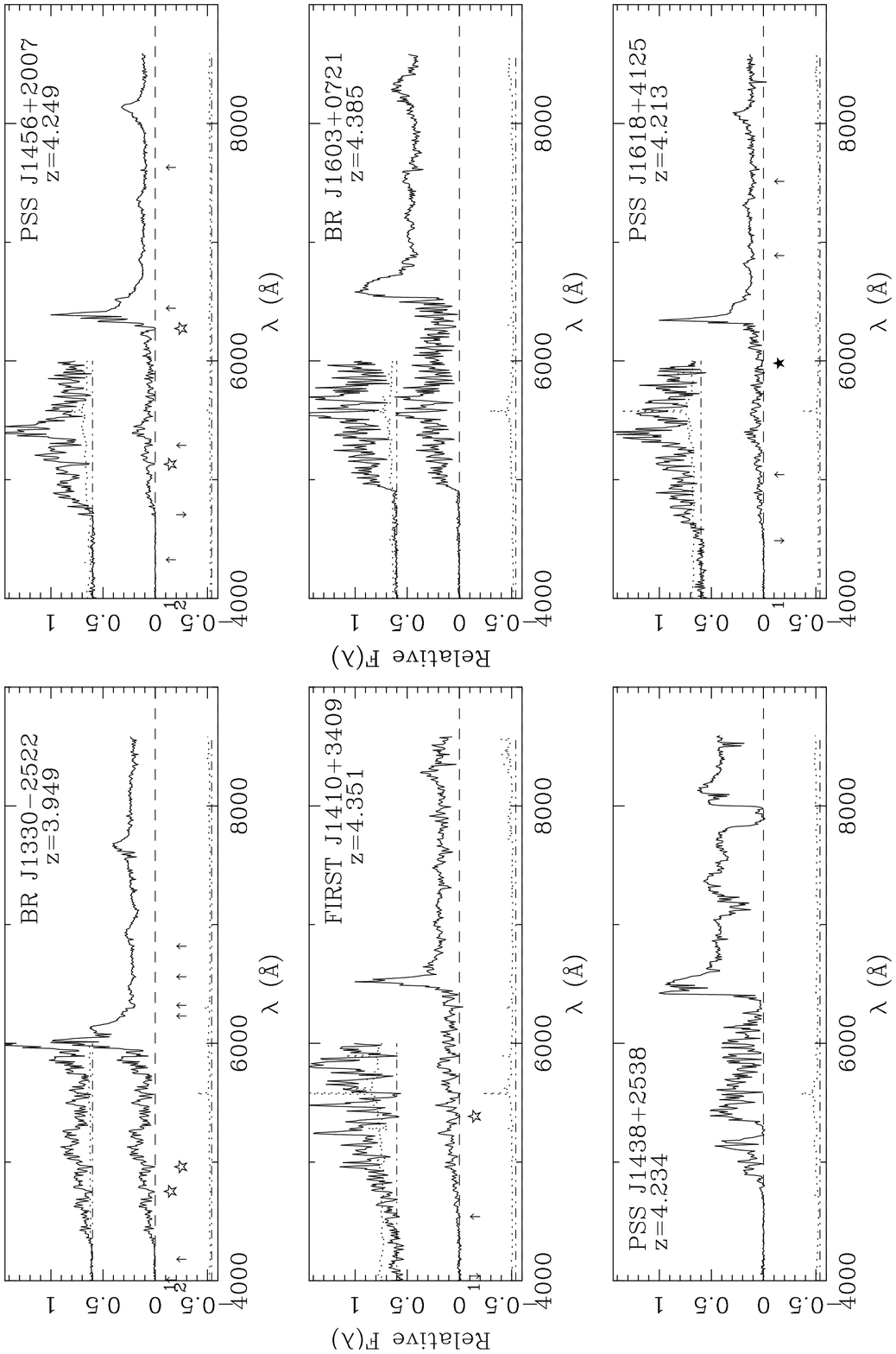}
\setcounter{figure}{0}
\figcaption{{\it continued}}
                      
\epsscale{.8}
\plotone{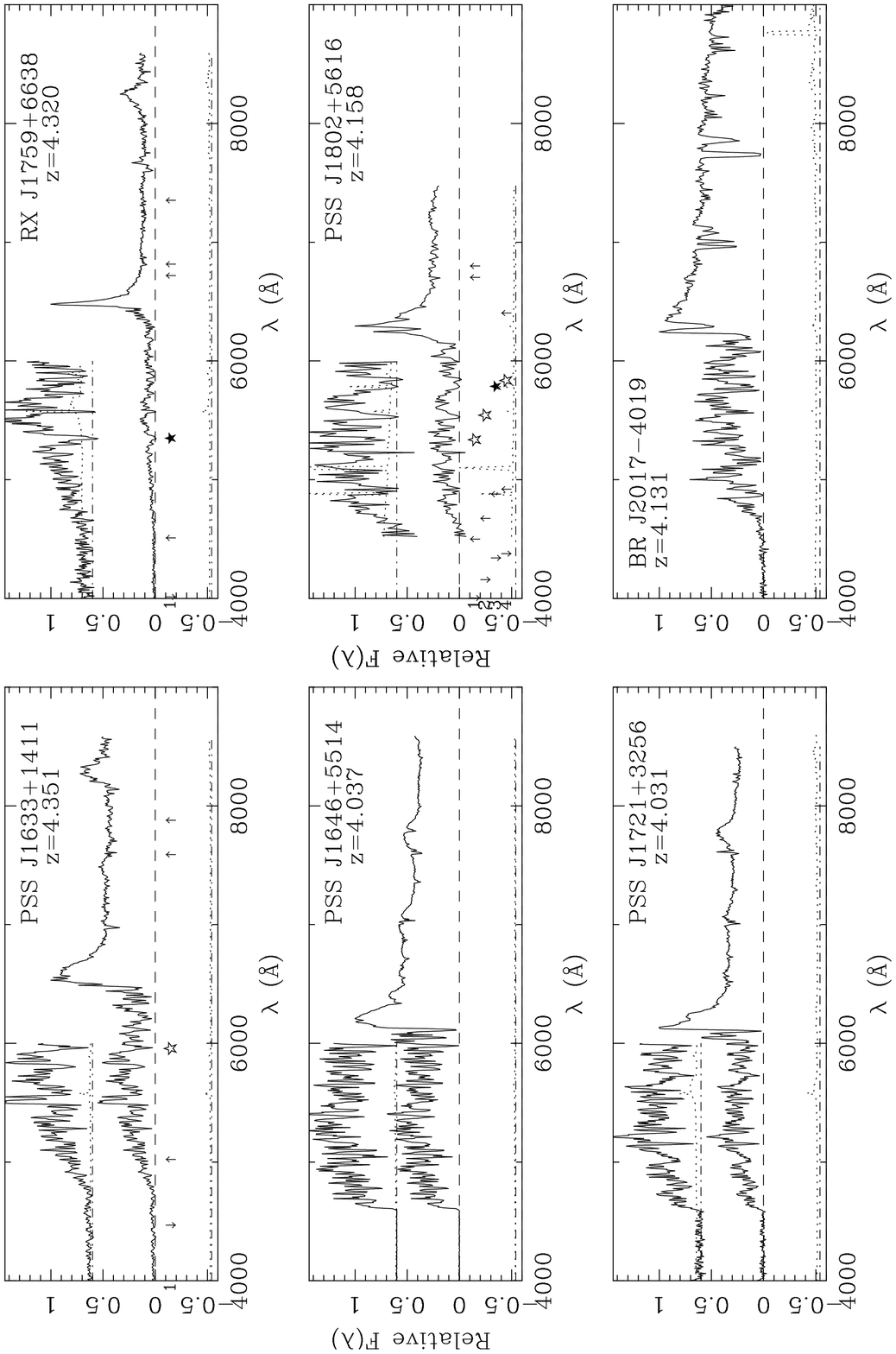}
\setcounter{figure}{0}
\figcaption{{\it continued}}
            
\epsscale{.8}
\plotone{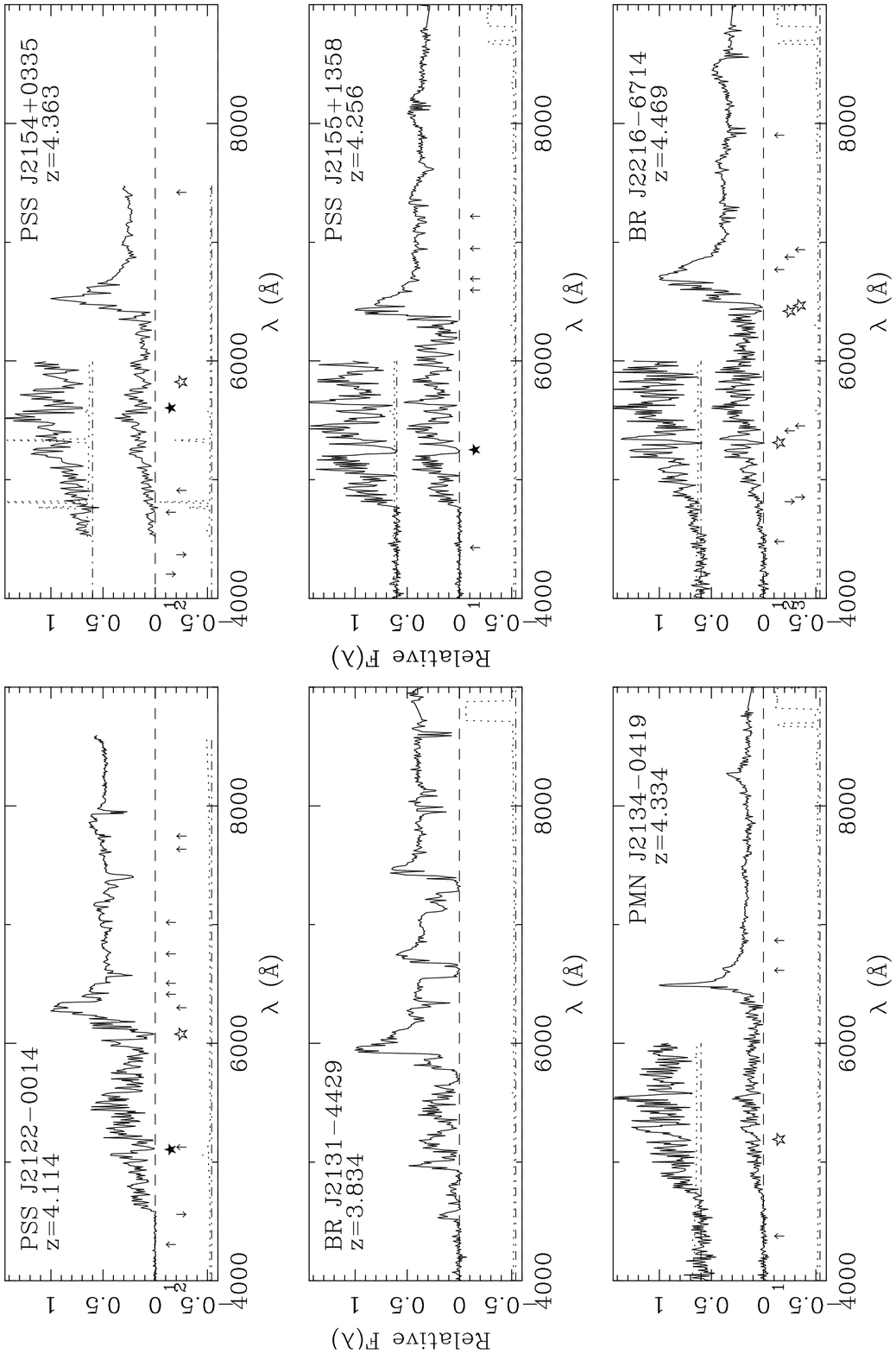}
\setcounter{figure}{0}
\figcaption{{\it continued}}
            
\epsscale{.8}
\plotone{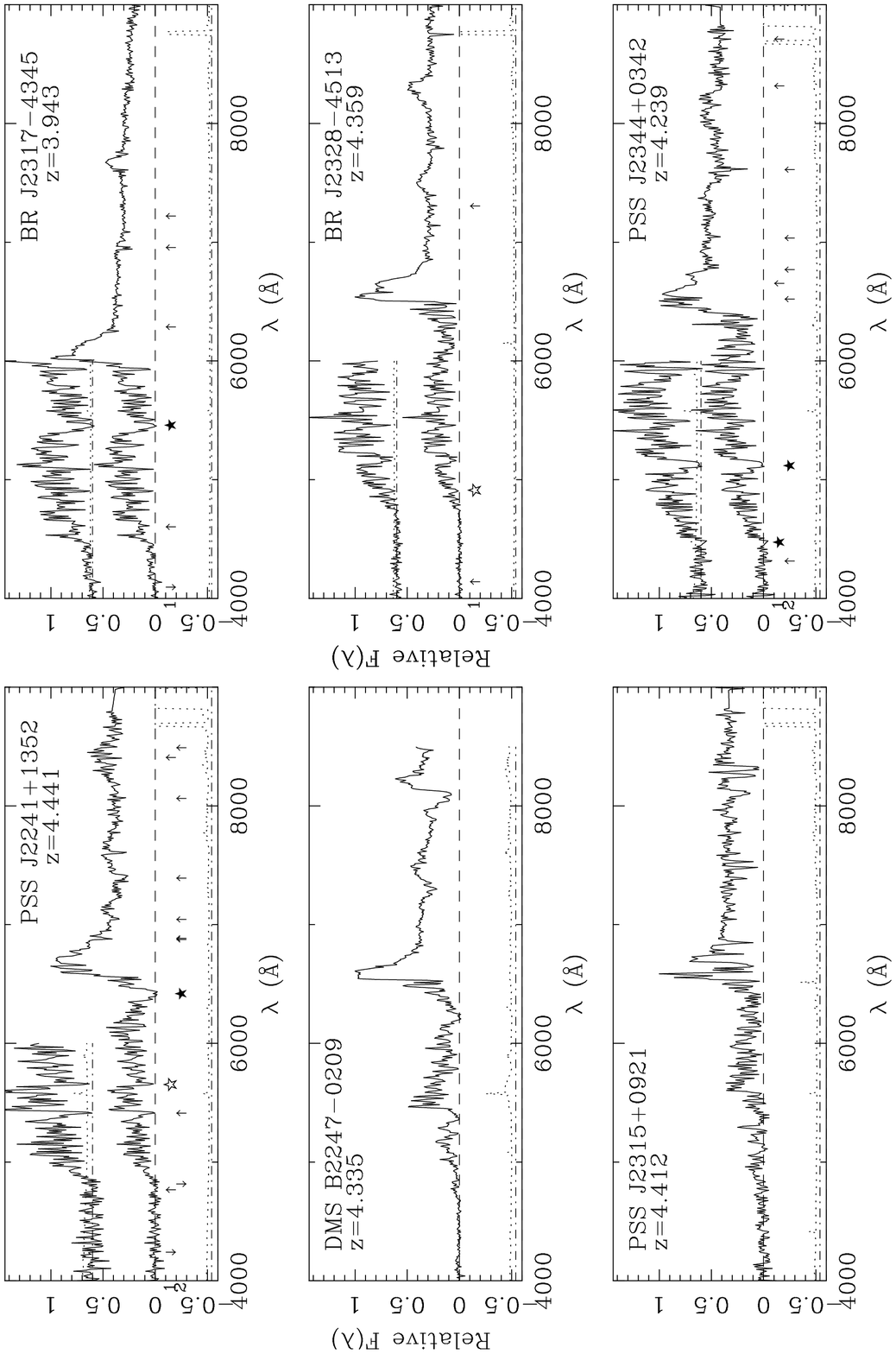}
\setcounter{figure}{0}
\figcaption{{\it continued}}
            
\epsscale{.6}
\plotone{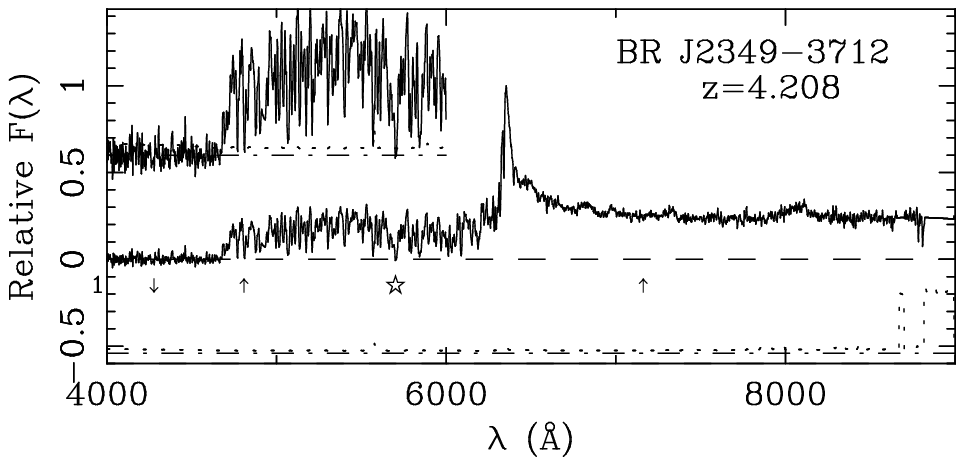} 
\setcounter{figure}{0}
\figcaption{{\it continued}}


\vspace{.5cm}
\epsscale{.5}
\plotone{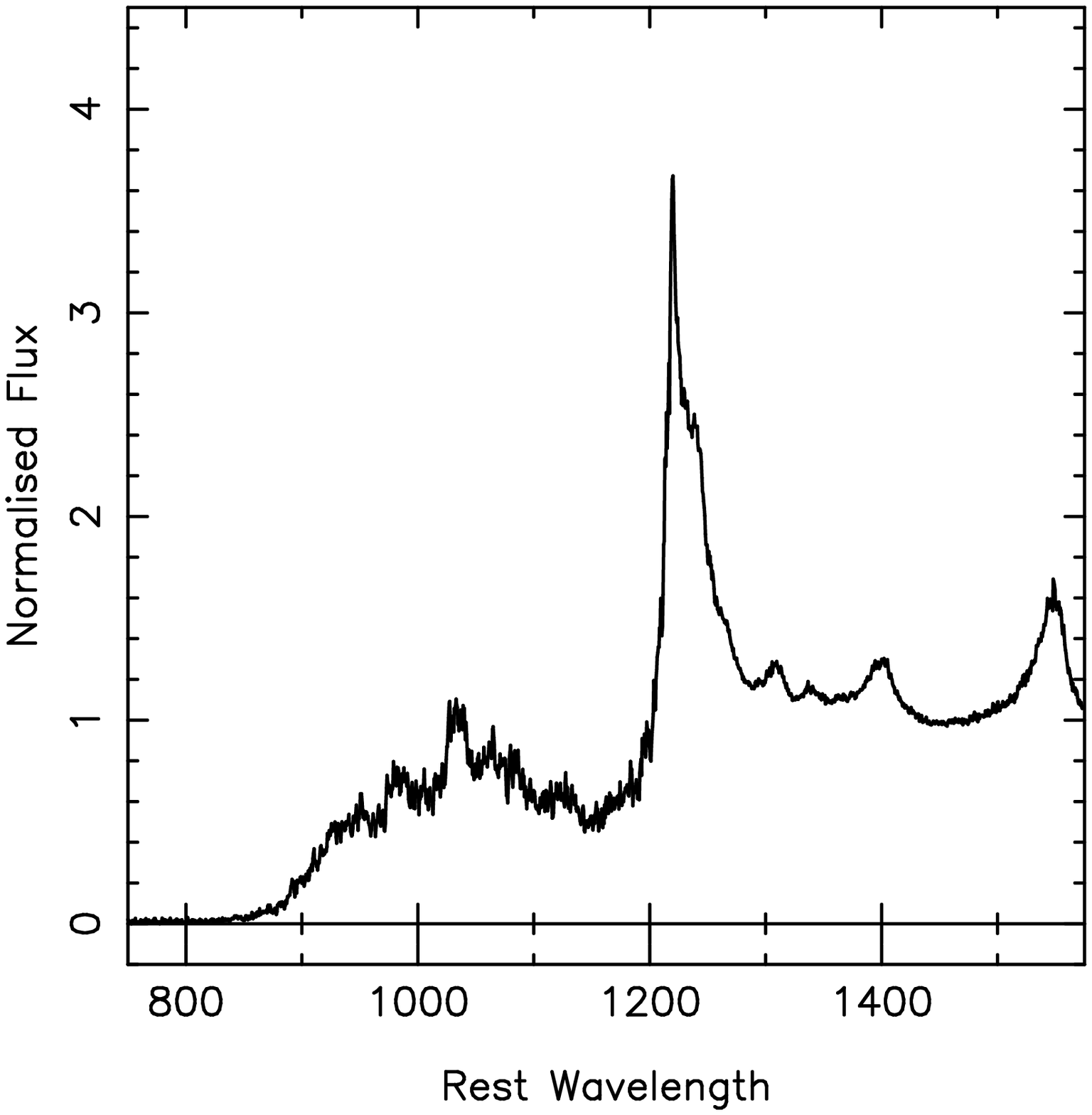} \figcaption{Median composite spectrum. This is
constructed by correcting each of the non-BAL spectra with enough
wavelength coverage to rest frame and normalizing the flux over a
region free of emission features (1420-1470 \AA).
\label{f_medspec}}

\vspace{.5cm}
\epsscale{.5}
\plotone{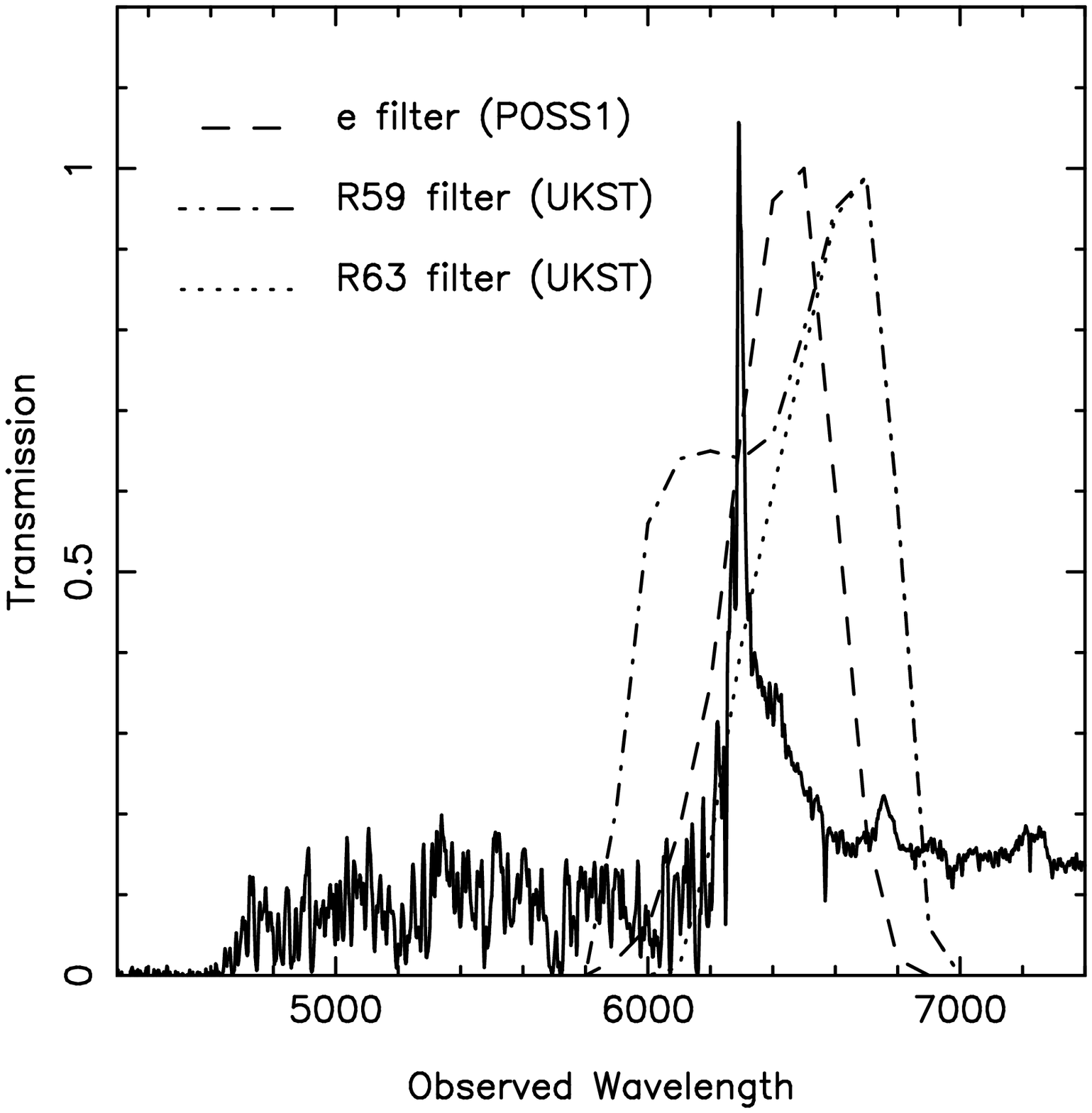} \figcaption{Filters used in various surveys. The R
filters used for the photographic plates scanned with the APM facility
are overplotted on the spectrum of z=4.172 BR J0529$-$3552. The ``e''
filter was used for the POSS1 survey and the ``R59'' (R) and ``R63''
(OR) were used for the UKST survey.
\label{f_filters}}

\vspace{.5cm}
\epsscale{1.}
\plottwo{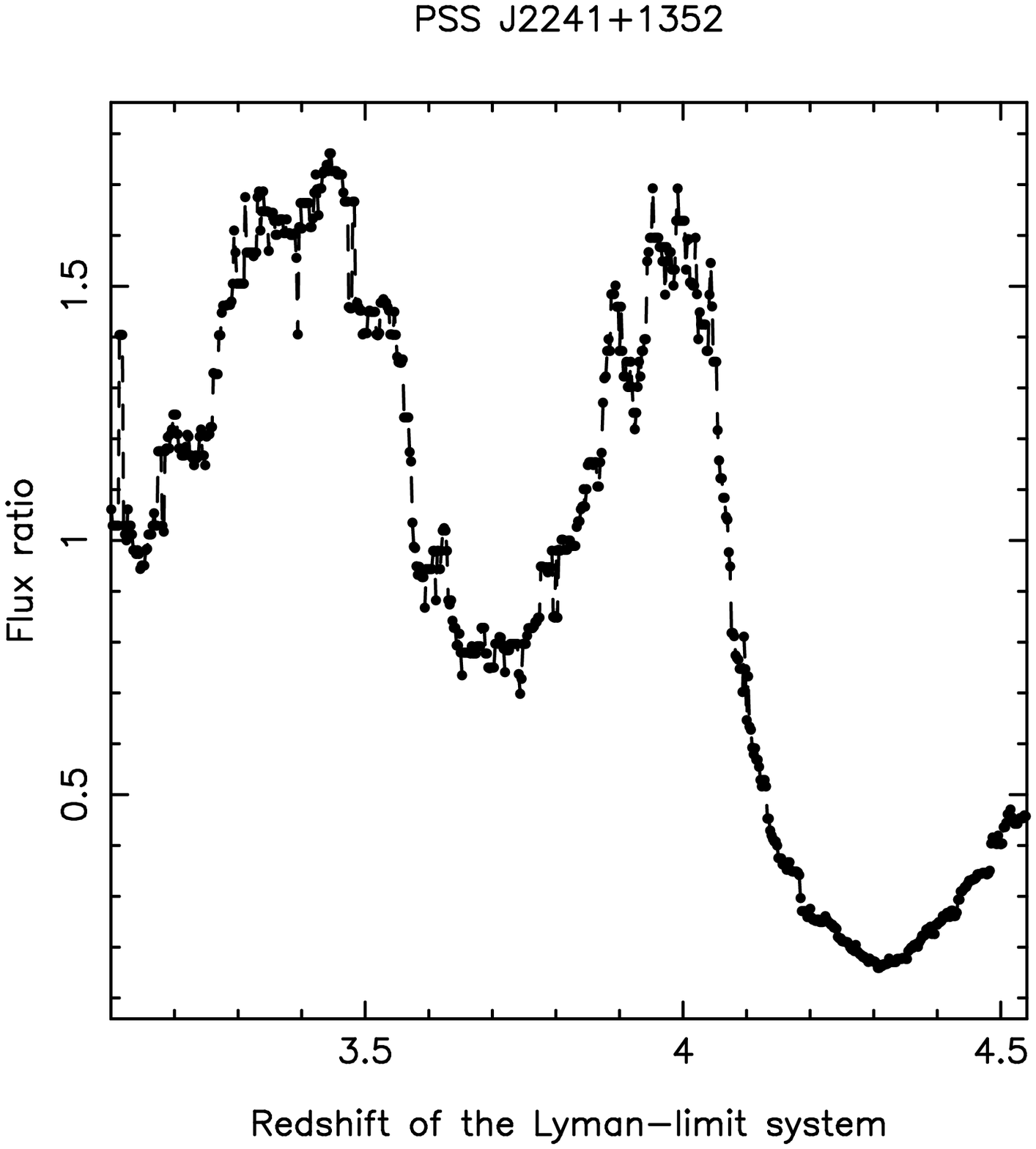}{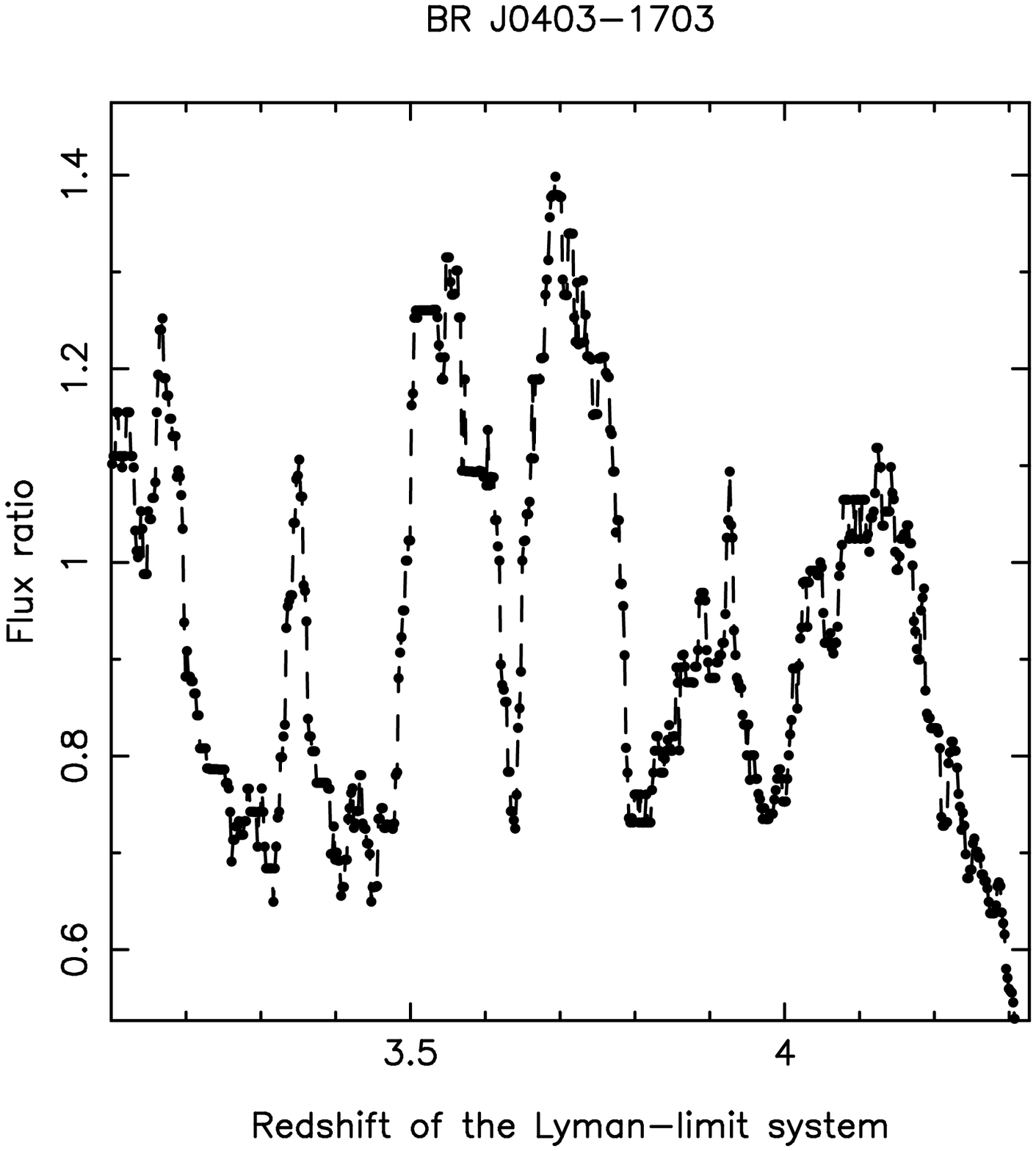} \figcaption{LLS
detection. Examples of flux ratio above and below the putative
Lyman-limit system redshifts. The plot for quasar PSS J2241$+$1352
(left panel) indicates a LLS with redshift $z = 4.31$ while the plot
for quasar BR J0403$-$1703 (right panel) does not show the presence of
a LLS.
\label{f_lls}}

\vspace{.5cm}
\plotone{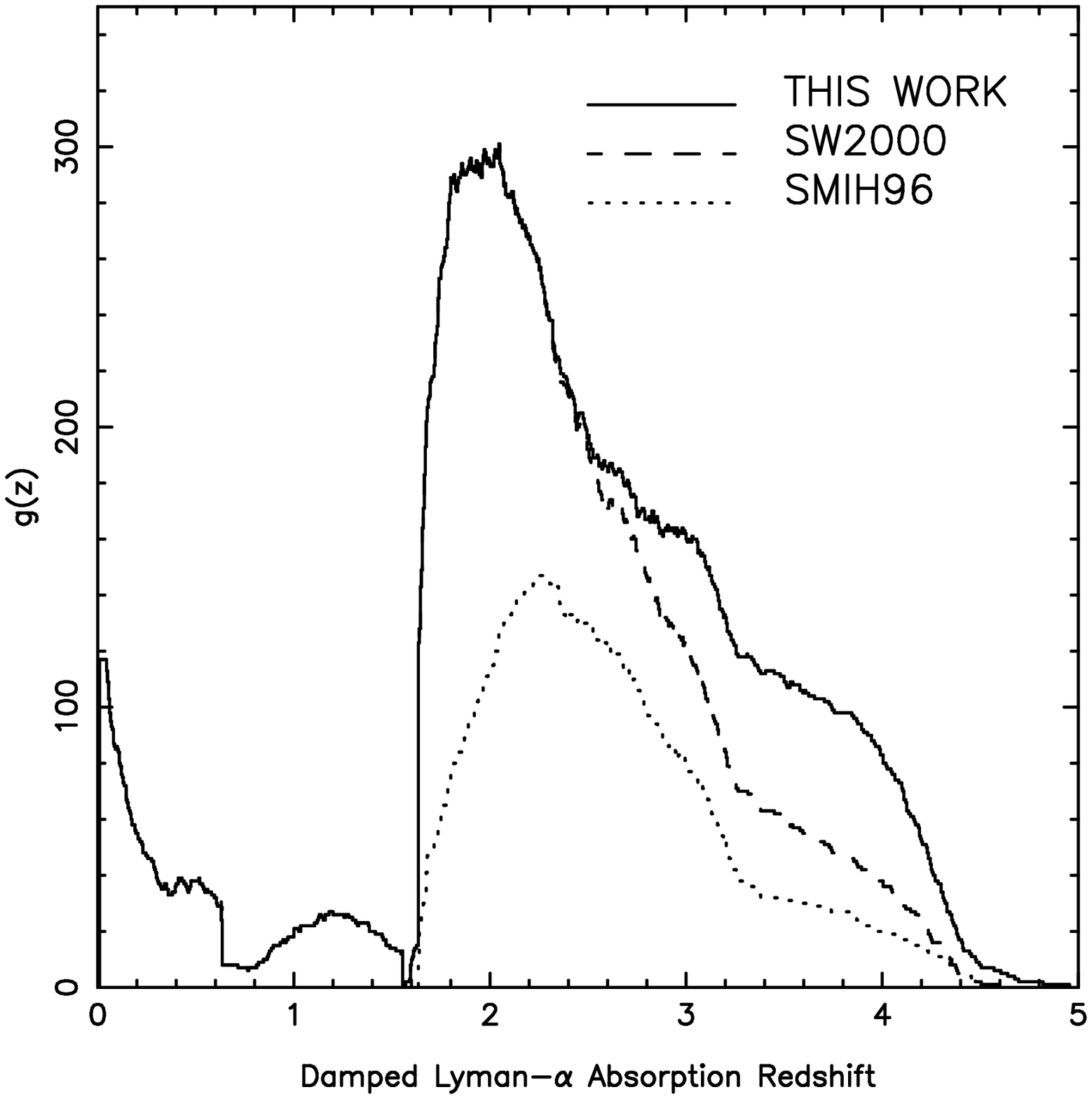} \figcaption{Survey sensitivity function. The g(z)
function shows the cumulative number of lines of sight along which a
DLA system could be detected if there was one. SW00 and SMIH96 are
surveys undertaken by Storrie-Lombardi \& Wolfe 2000 and
Storrie-Lombardi \etal\ 1996b, respectively. Our new observations more
than doubles the redshift path surveyed at $z\geq3.5$.
\label{f_g(z)}}

\vspace{.5cm}
\epsscale{.9}
\plotone{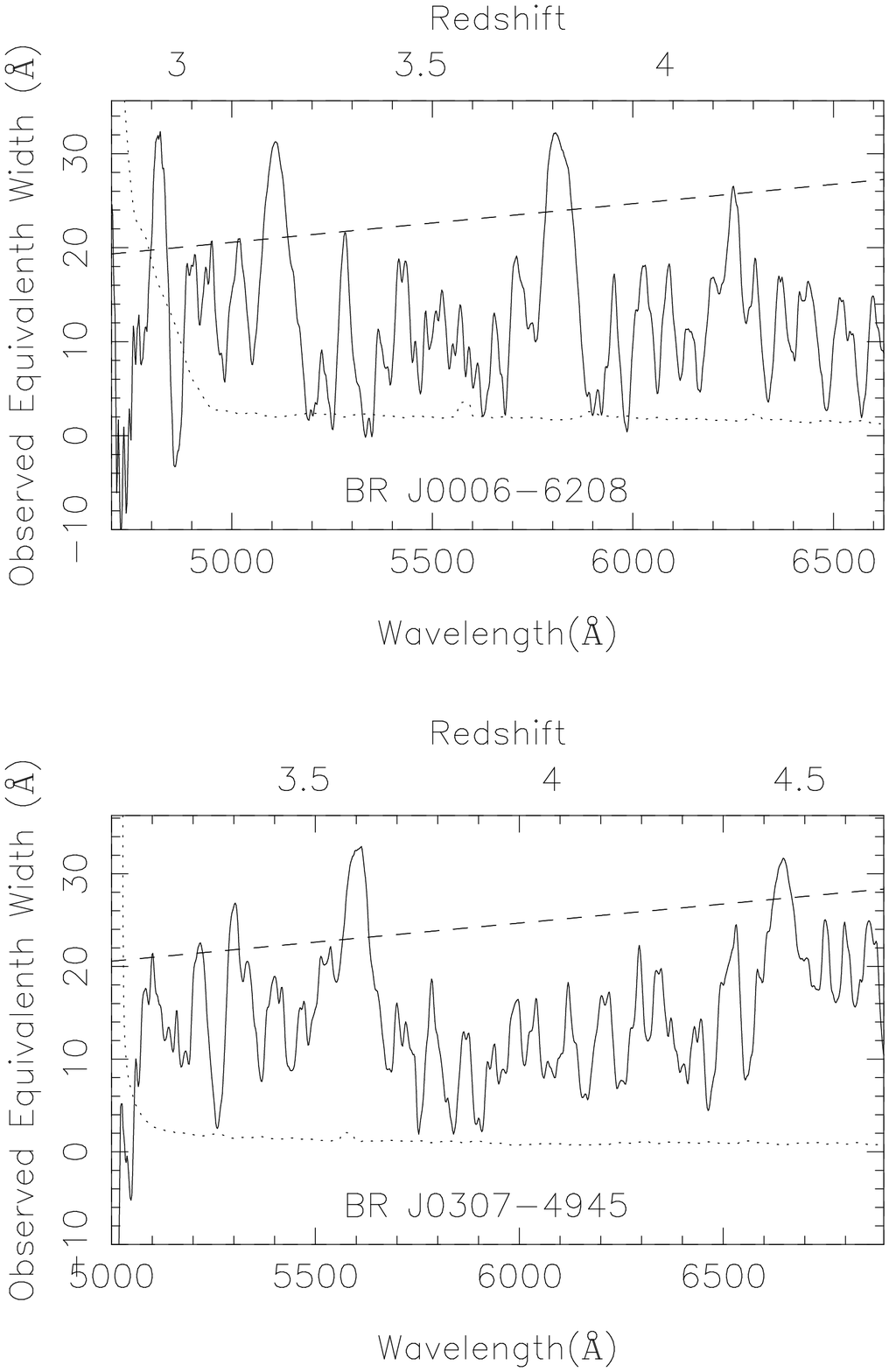} \figcaption{DLA detection. The figure shows
two examples of the output from the algorithm that detects damped
{\lya} absorption candidates.  The spectrum equivalent width bins are
shown as a solid line, the error equivalent width bins are shown as a
dotted line, and the dashed line shows the observed equivalent width
of a 5 \AA\ rest equivalent width line at the redshifts shown along the
top axis.  The lower axis shows the wavelength scale.  The minimum
redshift ($z_{min}$ in table~\ref{t_dlacand}) to which we can survey
for damped candidates is determined by the point where the error line
(dotted) crosses the 5 \AA\ rest equivalent width threshold (dashed
line).  The places where the spectrum array (solid line) goes above
the dashed line threshold are the wavelengths at which we measure the
aequivalent width of the lines in the original spectrum. The upper panel
shows four potential absorbers in BR~J0006-6208 and the lower panel
shows 5 potential absorbers in BR~J0307-4945.
\label{f_finddla}}

\vspace{.5cm}
\epsscale{1.2}
\plotone{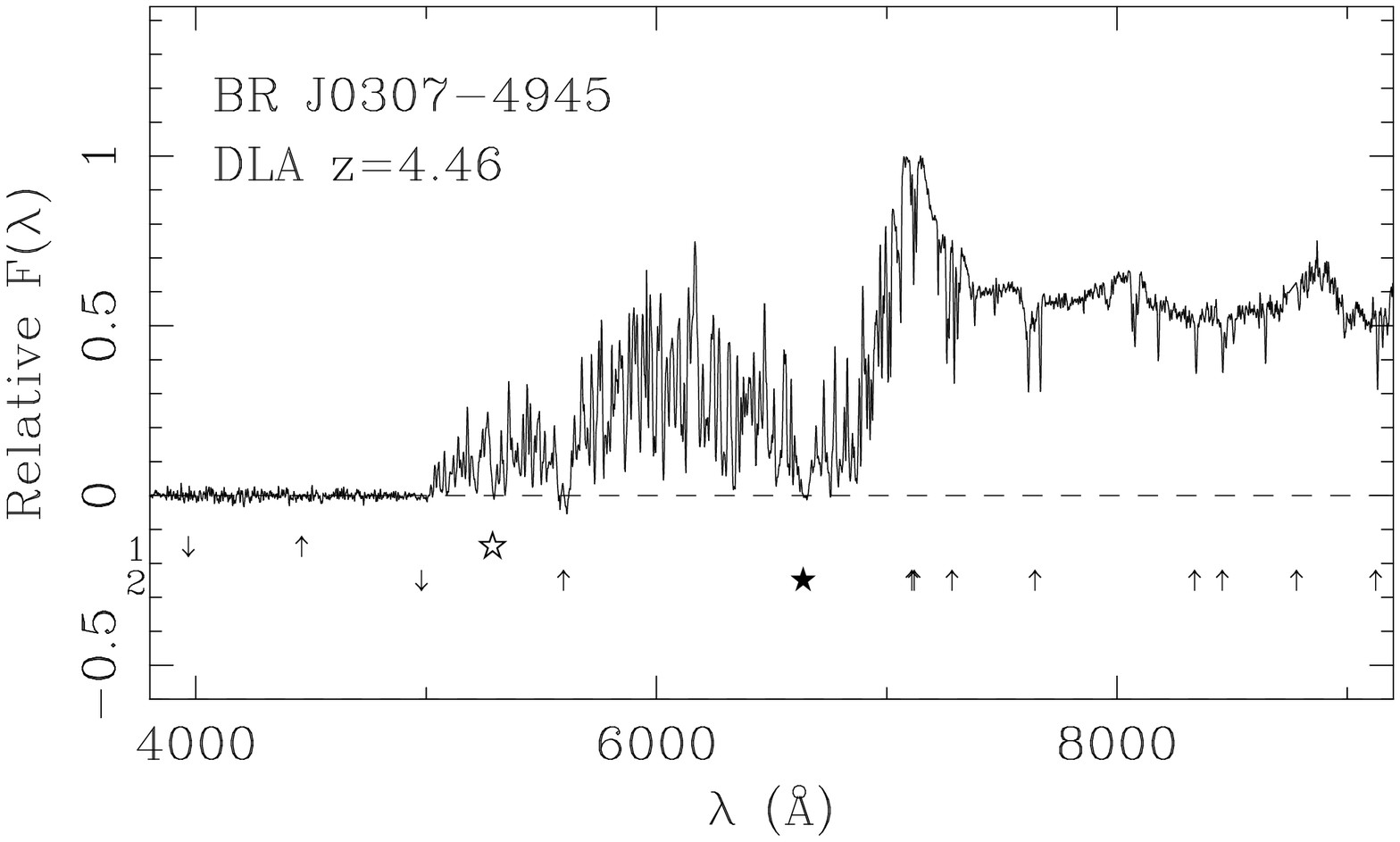} \figcaption{Example of DLA candidates. The spectrum of
quasar BR~J0307$-$4945 with DLA candidates marked at z=4.46 and z=3.35
is shown. The z=4.46 absorber is the highest redshift damped absorber
currently detected. The notations are as in Figure~\ref{f_spec}. Many
metal lines are observed at z=4.46 but no metals are detected at
z=3.35.  The higher redshift DLA has been studied in detail with
higher-resolution observations undertaken with the UVES spectrograph
(Dessauges-Zavadsky \etal\ 2000).
\label{f_q0307dla}}

\end{document}